\documentclass[12pt]{article}

\usepackage{thumbpdf,lmodern}
\usepackage[margin=1in]{geometry}

\usepackage{listings}
\usepackage{xcolor}
\usepackage{hyperref}
\usepackage{amssymb}
\usepackage{graphicx} 
\usepackage{fancyvrb}
\usepackage{stata}
\usepackage{rotating} 
\usepackage{amsmath}
\usepackage{algorithm}
\usepackage{algpseudocode}

\usepackage{booktabs} 
\usepackage{array} 
\usepackage{paralist} 
\usepackage{verbatim} 
\usepackage{subfig} 

\usepackage{caption}

\usepackage{listings}
\usepackage{color}
\newenvironment{CodeChunk}{}{}

\definecolor{codegreen}{rgb}{0,0.6,0}
\definecolor{codegray}{rgb}{0.5,0.5,0.5}
\definecolor{codepurple}{rgb}{0.58,0,0.82}
\definecolor{backcolour}{rgb}{0.95,0.95,0.92}

\lstdefinestyle{mystyle}{
    backgroundcolor=\color{backcolour},   
    commentstyle=\color{codegreen},
    keywordstyle=\color{magenta},
    numberstyle=\tiny\color{codegray},
    stringstyle=\color{codepurple},
    basicstyle=\ttfamily\footnotesize,
    breakatwhitespace=false,         
    breaklines=true,                 
    captionpos=b,                    
    keepspaces=true,                 
    numbers=left,                    
    numbersep=5pt,                  
    showspaces=false,                
    showstringspaces=false,
    showtabs=false,                  
    tabsize=2
}

\newtheorem{definition}{Definition}
\numberwithin{definition}{subsection} 

\DefineVerbatimEnvironment{Sinput}{Verbatim}{fontshape=sl}
\DefineVerbatimEnvironment{Soutput}{Verbatim}{}
\DefineVerbatimEnvironment{Scode}{Verbatim}{fontshape=sl}

\DefineVerbatimEnvironment{Code}{Verbatim}{}
\DefineVerbatimEnvironment{CodeInput}{Verbatim}{fontshape=sl}
\DefineVerbatimEnvironment{CodeOutput}{Verbatim}{}
\newcommand\code{\bgroup\@makeother\_\@makeother\~\@makeother\$\@codex}
\def\@codex#1{{\normalfont\ttfamily\hyphenchar\font=-1 #1}\egroup}

\usepackage{natbib}

\newcommand{\E}{E}

\newcommand{\Prob}{P}

\title{PySDTest: a Python/Stata Package for Stochastic Dominance Tests}
\author{Kyungho Lee\thanks{Email: \texttt{kyungho.lee@yale.edu} Department of Economics, Yale University} \and Yoon-Jae Whang\thanks{Email: \texttt{whang@snu.ac.kr}. Department of Economics, Seoul National University.  This work was supported by the 2021 Development Fund of the Department of Economics at Seoul National University.}}
\date{\today}

\begin{document}

\maketitle

\begin{abstract}
We introduce PySDTest, a Python/Stata package for statistical tests of stochastic dominance. PySDTest implements various testing procedures such as \cite{barrett2003consistent}, \cite{linton2005consistent}, \cite{linton2010improved}, and \cite{donald2016improving}, along with their extensions. Users can flexibly combine several resampling methods and test statistics, including the numerical delta method \citep{dumbgen1993nondifferentiable, hong2018numerical, fang2019inference}. The package allows for testing advanced hypotheses on stochastic dominance relations, such as stochastic maximality among multiple prospects. We first provide an overview of the concepts of stochastic dominance and testing methods. Then, we offer practical guidance for using the package and the Stata command \texttt{pysdtest}. We apply PySDTest to investigate the portfolio choice problem between the daily returns of Bitcoin and the S\&P 500 index as an empirical illustration. Our findings indicate that the S\&P 500 index returns second-order stochastically dominate the Bitcoin returns. \\

\textbf{Keywords}: \texttt{pysdtest}, stochastic dominance, statistical testing, resampling, bitcoin
\end{abstract}

\newpage

\section{Introduction}

Stochastic dominance (SD) is an ordering rule of distribution functions, originally suggested by \citet{hadar1969rules} and \cite{hanoch1969efficiency}. It is based on the expected utility paradigm and gives a uniform ordering of prospects that does not depend on a specific preference structure (or utility function) of a decision maker. Because of its generality and nonparametric feature, academics and practitioners in various areas, such as economics, finance, insurance, medicine and statistics, have regarded SD ordering as a fundamental concept of decision-making. They have empirically examined SD relations in diverse areas. Examples include the portfolio choice problem \citep{post2003empirical,bali2009bonds}, income inequality and welfare analysis \citep{amin2003does, anderson2003poverty}, financial market efficiency \citep{cho2007there}, auction bids \citep{de2003empirical}, firm productivity and export decision \citep{delgado2002firm}, obesity inequality \citep{pak2016measuring}, and agricultural productivity \citep{langyintuo2005yield} among many others. The number of applications has been growing along with recent advances in inference methods for testing SD hypotheses.

There exists a big literature for statistical tests of SD; see \citet{whang2019econometric} for an extensive survey.  \citet{mcfadden1989testing} pioneered statistical tests of SD, proposing a Kolmogorov-Smirnov type test for the first and second order SD. Other early works are \citet{klecan1991robust, kaur1994testing, anderson1996nonparametric, davidson2000statistical}.  \citet[BD]{barrett2003consistent} introduce resampling procedures for consistent testing of arbitrary SD order between two distributions under i.i.d. and mutually independent sampling schemes. \citet*[LMW]{linton2005consistent} propose a consistent subsampling method which is valid under general sampling schemes, including serially dependent observations and mutually dependent prospects. The LMW test also allows for prospects that may depend on some unknown finite-dimensional parameters, enabling comparison between residuals after controlling for observed confounders. \citet*[LSW]{linton2010improved} introduce the \textit{contact set approach} to improve the power performance of SD tests and show large sample uniform validity of the test. Also, they allow the prospects to depend on unknown \textit{infinite} dimensional parameters. \citet[DH]{donald2016improving} provide an alternatvie method to enhance the performance by suggesting the \textit{selective recentering approach}. 
However, to our knowledge, comprehensive statistical software for conducting SD tests is not available. 

This paper introduces a Python package PySDTest and Stata command \texttt{pysdtest} for testing SD hypotheses, hopefully to help practitioners implement SD tests in a convenient and flexible way. The package implements various SD tests developed in the recent literature. Supported testing procedures include, but are not limited to, the least favorable case approach of BD, subsampling approach of LMW, contact set approach of LSW, and selective recentering method of DH. In addition, we include the numerical delta method (NDM) based on the results of \citet{dumbgen1993nondifferentiable}, \citet{fang2019inference}, and \citet{hong2018numerical} in a more general context to compute critical values, as the NDM may have computational advantages in some contexts. This is a novel result since there has been no practical software tailored to SD testing. We provide guidance on how to use our package in Python and Stata environments with practical examples.

PySDTest also allows for flexibly testing a complex and advanced null hypothesis based on SD relations. For example, practitioners can compare more than 2 prospects simultaneously. To be specific, given $K (\geq 3)$ prospects, they can perform a joint test whether there exists at least one prospect stochastically dominating some of the others, which is called \textit{stochastic non-maximality} among multiple prospects \citep{klecan1991robust}. If they reject the hypothesis, then they may confer that the set of prospects is \textit{stochastically maximal}, in the sense that no prospect in the set is dominated by any of the others. This practice is possible because the package offers additional features such as diverse resampling methods and functional type of test statistics. We describe them in detail with an illustrative example.

As an empirical illustration, we apply PySDTest to the portfolio choice problem in the cryptocurrency and stock markets. We investigate the SD relation between the daily returns of Bitcoin price and S\&P 500 index. Our testing result shows that the S\&P 500 returns second order stochastically dominate the Bitcoin returns. This implies that a risk-averse economic agent with a monotone increasing utility function would prefer S\&P 500 index to Bitcoin as a financial asset. 

This paper is organized as follows. In Section 2, we introduce the concepts of SD and their testing methods. In Section 3, we provide a full description of the Python package PySDTest and Stata command \texttt{pysdtest}. In Section 4, we apply PySDTest to compare the daily return distributions of Bitcoin price and S\&P 500 index. In Section 5, we conclude the paper.

\section{Concepts and Tests of Stochastic Dominance}
\subsection{Concepts of SD}

Let $X_{1}$ and $X_{2}$ be two (continuous) random variables with cumulative distribution functions (CDF) $F_{1}$ and $F_{2}$, respectively. Let $Q_{k} = \inf \{x: F_{k}(x) \geq \tau \}$ denote the quantile function of $X_{k}$ for $k=1,2$, respectively. In addition, let $\mathcal{U}_{1}$ be the class of all monotone increasing utility (or social welfare) functions. If the functions are differentiable, then $\mathcal{U}_1$ can be written as:

\begin{equation*}
    \mathcal{U}_{1} = \{u(\cdot): u' \geq 0 \}
\end{equation*}

\begin{definition}[FSD]
The random variable $X_{1}$ is said to first-order stochastically dominates the random variable $X_{2}$, denoted by $F_{1}\succeq_{1} F_{2}$ if any of the following equivalent conditions holds: (1) $F_{1}(x) \leq F_{2}(x)$ for all $x \in \mathbb{R}$; (2) $E[u(X_{1})] \geq E[u(X_{2})]$ for all $u \in \mathcal{U}_{1}$; and (3) $Q_{1}(\tau) \geq Q_{2}(\tau)$ for all $\tau \in [0,1]$.
\end{definition}

This definition can be understood by considering an investor's portfolio choice problem. Let the random variables be returns of some financial assets. The condition $F_{1}(x) \leq F_{2}(x)$ for all $x \in \mathbb{R}$ can be denoted as:

\begin{equation}
    P(X_{1} > x) \geq P(X_{2} > x) \text{ for all } x \in \mathbb{R}
\end{equation}
which means, for all $x \in \mathbb{R}$, the probability of attaining returns larger than $x$ is larger under $F_{1}$ than $F_{2}$. An investor with a monotone increasing utility function would therefore prefer $F_{1}$ to $F_{2}$. If the two CDFs intersect, however, FSD does not hold. This means that there exist some $u$ and $v$ both in  $\mathcal{U}_{1}$ such that $E[u(X_{1})] > E[u(X_{2})]$, but $E[v(X_{1})] < E[v(X_{2})]$.

The second-order stochastic dominance can be defined in a similar manner. Let $\mathcal{U}_{2}$ be the class of all monotone increasing and concave (utility or social welfare) functions. If functions are differentiable, then we can write the class $\mathcal{U}_2$ as:

\begin{equation*}
    \mathcal{U}_{2} = \{u(\cdot): u' \geq 0, u'' \leq 0 \}.
\end{equation*}

\begin{definition}[SSD]
The random variable $X_{1}$ is said to second-order stochastically dominates the random variable $X_{2}$, denoted by $F_{1}\succeq_{2} F_{2}$ if any of the following equivalent conditions holds: (1) $\int_{-\infty}^{x}F_{1}(t)dx \leq \int_{-\infty}^{x}F_{2}(t)dx $ for all $x \in \mathbb{R}$; (2) $E[u(X_{1})] \geq E[u(X_{2})]$ for all $u \in \mathcal{U}_{2}$; and (3) $\int_{0}^{\tau}Q_{1}(p)dp \geq \int_{0}^{\tau}Q_{2}(p)dp$ for all $\tau \in [0,1]$.
\end{definition}

Consider the portfolio choice problem of an investor again. The concavity of utility function reflects risk-aversion of the investor. If $X_{1} \succeq_{2} X_{2}$, which represents $E[u(X_{1})] \geq E[u(X_{2})]$ for all $u \in \mathcal{U}_{2}$, the risk-averse investor would prefer $X_{1}$ to $X_{2}$. 

Higher-order SD relations also can be defined. Define \emph{integrated CDF} and \emph{integrated quantile function} as \citep{davidson2000statistical}: 

\begin{equation}
    F_{k}^{(s)}(x) = \begin{cases}
F_{k}(x) &\text{ for } s=1\\
\int_{-\infty}^{x}F_{k}^{(s-1)}(t)dt &\text{ for } s\geq 2
\end{cases}
\end{equation}

\begin{equation}
    Q_{k}^{(s)}(x) = \begin{cases}
Q_{k}(x) &\text{ for } s=1\\
\int_{0}^{x}Q_{k}^{(s-1)}(t)dt &\text{ for } s\geq 2
\end{cases}
\end{equation}
The higher SD order corresponds to the smaller class of utility (or social welfare) functions. That is, for $s \geq 1$, the class of utility functions is defined as:
\begin{equation*}
    \mathcal{U}_{s} = \{u(\cdot): u' \geq 0, u'' \leq 0,..., (-1)^{(s+1)}u^{(s)} \geq 0 \}.
\end{equation*}

\begin{definition}
The random variable $X_{1}$ is said to s-order stochastically dominates the random variable $X_{2}$, denoted by $F_{1}\succeq_{s} F_{2}$ if any of the following equivalent conditions holds: (1) $F_{1}^{(s)}(x) \leq F_{2}^{(s)}(x) $ for all $x \in \mathbb{R}$ and  $F_{1}^{(r)}(\infty) \leq F_{2}^{(r)}(\infty)$ for all $r=1,...,s-1$; (2) $E[u(X_{1})] \geq E[u(X_{2})]$ for all $u \in \mathcal{U}_{s}$; and (3) $Q_{1}^{(s)}(\tau) \geq Q_{2}^{(s)}(\tau)$ for all $\tau \in [0,1]$ and $Q_{1}^{(r)}(1) \geq Q_{2}^{(r)}(1)$ for all $r=1,...,s-1$.
\end{definition}

For example, the third-order stochastic dominance ($s=3$) reflects preference toward positive skewness of a distribution (\cite{whitmore1970third}, \cite{whitmore1978stochastic}, and \cite{levy2016stochastic}). That is, investors would prefer positively skewed distributions to negatively skewed distributions. 

\subsection{Tests of SD}

\subsubsection[Hypotheses of interest]{Hypothesis of interest}
For a given stochastic order $s$, consider the following hypotheses of interest:
\begin{equation} \label{eq:h0}
    H_{0}^{s}: F_{1}^{(s)}(x) \leq F_{2}^{(s)}(x) \text{ for all } x \text{ v.s. } H_{1}^{s}: F_{1}^{(s)}(x) > F_{2}^{(s)}(x) \text{ for some } x 
\end{equation} The null hypothesis means that random variable $X_1$ $s$-th order stochastically dominates $X_2$ and the alternative hypothesis is the negation of the null hypothesis.\footnote{This type of hypotheses has mainly been considered  for statistical inference of the SD relations in the literature. There are two other types of hypotheses that are also considered. The first one is the null hypothesis of non-dominance and the alternative hypothesis of dominance. The other type is the null hypothesis of equivalent distributions and the alternative hypothesis of (strict) stochastic dominance. In this paper, we do not cover these hypotheses. For further discussions, see Section 2.1 of \cite{whang2019econometric}.}

\subsubsection[BD Test: Least Favorable Case Approach]{BD Test: Least Favorable Case Approach}

Let $\{X_{1,i} :  i = 1,2,...,N_1\}$ and $\{X_{2,i} :  i = 1,2,...,N_2\}$ be random samples from $F_1$ and $F_2$, respectively. We assume $N_1 = N_2 \equiv N$ for simplification.\footnote{PySDTest allows for the case with different sample sizes  $N_1 \neq N_2 $.} Let $\mathcal{X}$ denote the common compact support of $X_1$ and $X_2$. 

We first introduce and define frequently used notions in the SD testing literature. The integrated CDF, $F_k^{(s)}(\cdot)$, can be written through integration by parts as \citep{davidson2000statistical}:
\begin{equation} \label{eq:operator}
    F_k^{(s)}(x) \equiv \frac{1}{(s-1)!}\int_{0}^{x}(x-t)^{s-1}dF_k(t) =  \frac{1}{(s-1)!}\E\left[x - X_k\right]_{+}^{s-1},
\end{equation} where $\left[\cdot \right]_{+} = \max\{\cdot,0\}$. The natural empirical counterpart, \emph{empirical integrated CDF}, of \eqref{eq:operator} is given by
\begin{equation} \label{eq:empirical-opreator}
    \Bar{F}_k^{(s)}(x) = \frac{1}{N(s-1)!}\sum_{i=1}^{N}\left[x - X_{k,i}\right]_{+}^{s-1}.
\end{equation} Define the difference between integrated CDFs and empirical integrated CDFs as:
\begin{align}
    D_{k,l}^{(s)}(x) = F_k^{(s)}(x) - F_l^{(s)}(x) \label{eq:D-kl}\\
    \Bar{D}_{k,l}^{(s)}(x) = \Bar{F}_k^{(s)}(x) - \Bar{F}_l^{(s)}(x).  \label{eq:bar-D-kl}
\end{align}

BD propose the supremum (or Kolmogorov-Smirnov) type test statistic

\begin{equation*} \label{eq:BD-test}
    BD_{N} = \sqrt{N}\sup_{x \in \mathcal{X}} \big( \Bar{D}_{1,2}^{(s)}(x)\big)
\end{equation*} to test the hypothesis \eqref{eq:h0}. Note that we can write \eqref{eq:operator} as:
\begin{equation} \label{eq:linear-op}
    \mathcal{J}_{s}(x,F_k) := F_k^{(s)}(x) = \frac{1}{(s-1)!}\int_{0}^{x}(x-t)^{s-1}dF_k(t),
\end{equation} where $\mathcal{J}_s(x,F_k)$ is a linear and continuous functional with respect to $F_k$. Under i.i.d. random sampling and compact support $\mathcal{X}$, the following weak convergence result holds:
\begin{equation} \label{eq:weak-conv-Fs}
    \sqrt{N}(\Bar{F}_{k}^{(s)}(\cdot) - F_{k}^{(s)}(\cdot)) \Rightarrow \mathcal{J}_s(\cdot, \mathcal{B}_k \circ F_k) \text{ for } k=1,2
\end{equation} as $N\rightarrow \infty$, where $\mathcal{B}_1$ and $\mathcal{B}_2$ are independent standard Brownian bridges and $\Rightarrow$ denotes the weak convergence. This result comes from applying Donsker theorem to the case $s=1$  and continuous mapping theorem for the case $s \geq 2$. The limit process, $\mathcal{J}_s(\cdot, \mathcal{B}_k \circ F_k)$, is a Gaussian process with the covariance function given by
\begin{equation*}
    C_s(x_1,x_2;F_k) = \E \mathcal{J}_s(x_1,  \mathcal{B}_k \circ F_k )\mathcal{J}_s(x_2,  \mathcal{B}_k \circ F_k ).
\end{equation*}
Define 
\begin{align*}
    \hat{T}_s(\cdot) &:= \sqrt{N}\left[(\Bar{F}^{(s)}_1(\cdot) - F^{(s)}_1(\cdot)) - (\Bar{F}^{(s)}_2(\cdot) - F^{(s)}_2(\cdot))\right]\\
    \nu_s(\cdot) &:= \mathcal{J}_s(\cdot,  \mathcal{B}_1 \circ F_1 ) - \mathcal{J}_s(\cdot,  \mathcal{B}_2 \circ F_2 ).
\end{align*} Equation \eqref{eq:weak-conv-Fs} implies 
\begin{equation} \label{eq:weak-conv-Ts}
\hat{T}_s(\cdot) \Rightarrow  \nu_s(\cdot).    
\end{equation}
Then, under the null hypothesis, we have the following result:
\begin{align*}
    BD_N &= \sup_{x \in \mathcal{X}}\left[ \hat{T}_s(x) + D_{1,2}^{(s)}(x) \right] \nonumber \\
    &\leq \sup_{x \in \mathcal{X}} \hat{T}_s(x) \text{ under }H_0 \nonumber
    \\&\Rightarrow \sup_{x\in \mathcal{X}}\nu_s(x) \nonumber \text{ by \eqref{eq:weak-conv-Ts} and the continuous mapping theorem.}
\end{align*} Let $c_{1-\alpha}^{BD}$ is the $(1-\alpha)$ quantile of $\sup_{x \in \mathcal{X}}\nu_s(x)$ for $\alpha \in [0,1]$. BD show the following results:
\begin{align}
    \limsup_{N \rightarrow \infty}&\Prob\left[BD_N \geq c_{1-\alpha}^{BD} \right] \leq \alpha \text{ under } H_0^s \label{eq:correct-size-BD} \\
    \lim_{N \rightarrow \infty}&\Prob\left[BD_N \geq c_{1-\alpha}^{BD} \right] = 1 \text{ under } H_1^s \nonumber
\end{align} The equality of \eqref{eq:correct-size-BD} holds when the least favorable case (LFC), $F_1 = F_2$, holds. Note that the limit distribution $\sup_{x \in \mathcal{X}}\nu_s(x)$ depends on the true CDF, $F_k$, hence it cannot be tabulated. 

BD propose three procedures to compute the critical values. All these methods impose the LFC restriction under the null hypothesis and try to mimic the distribution $\sup_{x\in \mathcal{X}} \nu_s(x)$. The first one is the multiplier method. It first generates two independent random samples $\{U_{k,i}:i=1,2,...,N\}$ from $N(0,1)$ for $k=1,2$. Then, simulate the process that mimics the $F_k-$Brownian bridge $\mathcal{B}_{k}\circ F_{k}(x)$:
\begin{equation*}
    \mathcal{B}_k^* \circ \Bar{F}_k(x) := \frac{1}{\sqrt{N}}\sum_{i=1}^{N} \left[1(X_{i,k}\leq x) - \Bar{F}_k(x)\right]U_{k,i} \text{ for } k=1,2.
\end{equation*} The multiplier method generates the following simulated process:
\begin{align*}
    \mathcal{J}_s(x, \mathcal{B}_{k}^* \circ \Bar{F}_k) &= \frac{1}{\sqrt{N}(s-1)!}\sum_{i=1}^{N}\left[(x - X_{k,i})^{s-1}1\left[X_{k,i} \leq x\right] - \frac{1}{N}\sum_{j=1}^{N}(x - X_{k,j})^{s-1}1\left[X_{k,j}  \leq x\right] \right]U_{k,i}
\end{align*} for $k=1,2$. Then, the critical value for the significance level $\alpha$ is defined as the $(1-\alpha)$ quantile of the following simulated distribution:
\begin{equation} \label{eq:mp-simul-BD}
    BD_{N,MP}^* :=\sup_{x\in \mathcal{X}}\{\mathcal{J}_s(x, \mathcal{B}_{1}^* \circ \Bar{F}_1) - \mathcal{J}_s(x, \mathcal{B}_{2}^* \circ \Bar{F}_2) \}.
\end{equation} Note that $BD_{N,MP}^*$ approximates the distribution of $\sup_{x \in \mathcal{X}} \nu_{s}(x)$ conditional on the original sample.

The second method is the pooled-sample bootstrap method. This method draws bootstrap samples $S_{k}^* = \{X_{k,i}^* : i=1,2,...,N\}$ for $k=1,2$ from the pooled original sample $\{X_{k,i}:k=1,2 \text{ and } i = 1,2,...,N\}$ and calculates the empirical CDFs by using the bootstrap samples as:
\begin{equation} \label{eq:emp-boot-cdf}
    \Bar{F}_k^*(x) = \frac{1}{N} \sum_{i=1}^{N}1(X_{k,i}^* \leq x) \text{ for }k=1,2.
\end{equation} Then, we compute the following distribution of the bootstrap test statistic:
\begin{equation} \label{eq:BD-pool}
    BD_{N, pool}^* :=\sup_{x\in \mathcal{X}}\{\mathcal{J}_s(x, \Bar{F}^*_1) - \mathcal{J}_s(x, \Bar{F}^*_2)\}.
\end{equation}The critical value for the significance level $\alpha$ is defined to be the $(1-\alpha)$ quantile of the $BD_{N,pool}^*$ distribution. Note that the LFC restriction is imposed by pooling the original samples. 

The third procedure is the recentered bootstrap method. This method draws bootstrap samples $S_{1}^*$ and $S_{2}^*$ from $\{X_{1,i}:i=1,2,...,N\}$ and $\{X_{2,i}:i=1,2,...,N\}$, respectively. Then, the empirical CDFs are calculated using each bootstrap sample as in \eqref{eq:emp-boot-cdf}. Define
\begin{equation} \label{eq:BD-recenter}
    BD_{N, recenter}^* :=  \sup_{x\in \mathcal{X}}\{(\mathcal{J}_s(x, \Bar{F}^*_1) - \mathcal{J}_s(x, \Bar{F}_1)) - ( \mathcal{J}_s(x, \Bar{F}^*_2) - \mathcal{J}_s(x, \Bar{F}_2))\},
\end{equation} where the LFC restriction is imposed by recentering the bootstrap statistic $\mathcal{J}_s(x, \Bar{F}^*_k)$ by subtracting its mean $\mathcal{J}_s(x, \Bar{F}_k)$ for $k=1,2$. The critical values are defined to be the $(1-\alpha)$ quantile of the bootstrap distribution $BD_{N,recenter}^*$.

\subsubsection[LMW Test: Subsampling Apporach for General Sampling Scheme]{LMW Test: Subsampling Apporach for General Sampling Scheme}

For a given set of $K$ prospects, LMW considered the following hypotheses:
\begin{align}
    H_{0} &: \min_{k \neq l}\{D_{k,l}^{(s)}(x) \} \leq 0 \text{ for all } x \in \mathcal{X} \label{eq:h0-lmw} \\
    H_{1} &: \min_{k \neq l}\{D_{k,l}^{(s)}(x) \} > 0 \text{ for some } x \in \mathcal{X}, \label{eq:h1-lmw}
\end{align} where $D_{k,l}^{(s)}(x)$ is defined as in \eqref{eq:D-kl} and $\mathcal{X}$ is the union of the supports of $X_k$ for $k=1,2,...,K$. The null hypothesis $H_0$ is the hypothesis of \textit{stochastic non-maximality} (\cite{klecan1991robust}). It says that there exists at least one prospect that $s$-th order stochastically dominates some of the others in the set of the $K$-prospects. The alternative hypothesis is that of \textit{stochastic maximality}, that is  there does not exist any prospect in the set which $s$-th order stochastic dominates any of the others.

LMW propose a supremum type test for testing \eqref{eq:h0-lmw} with the following features: (i) SD of \textit{arbitrary} order, (ii) Subsampling critical values allowing for \textit{general sampling scheme} such as serial dependence of observations and arbitrary dependence amongst prospects at a specific time, (iii) Prospects may depend on finite dimensional parameters, enabling \textit{conditional ranking}. For instance, the test can be applied to residuals from regressions to control for systematic factors. 

Let $X_{k,i}(\theta)$ be the prospect that may depend on a finite dimensional parameter $\theta$ for $k=1,2...,K$. If we consider the residual from the linear regression, we may take $X_{k,i}(\theta) = Y_{k,i} - Z_{k,i}^\intercal \theta$. Define
\begin{align*}
    F_{k}^{(1)}(x,\theta) &:= F_{k}(x,\theta) = P(X_{k,i}(\theta) \leq x) \\
    F_{k}^{(s)}(x,\theta) &:= \int_{-\infty}^{x}F_{k}^{(s-1)}(t,\theta)dt \text{ for } s\geq 2.
\end{align*} Likewise, we define $\Bar{F}_k^{(s)}(x,\theta)$ as the sample analogue of $F_{k}^{(s)}(x,\theta)$ for $s \geq 1$. 

Let $F_{k}^{(s)}(x) := F_{k}^{(s)}(x,\theta_{k,0})$ and $\Bar{F}_{k}^{(s)}(x) := \Bar{F}_{k}^{(s)}(x,\theta_{k,0})$, where $\theta_{k,0}$ denotes the true value, and define
\begin{equation*} 
    d_s := \min_{k \neq l}\sup_{x \in \mathcal{X}}\left[D_{k,l}^{(s)}(x) \right].
\end{equation*} We can represent the hypotheses \eqref{eq:h0-lmw} and \eqref{eq:h1-lmw} as:
\begin{equation*} \label{eq:h0-h1-ds}
    H_0 : d_s \leq 0 \text{ v.s. } H_1 : d_s > 0.
\end{equation*} LMW use the sample analogue of $d_s$ as the basis of their test statistic:
\begin{equation*}
    LMW_{N} = \sqrt{N}\min_{k\neq l}\sup_{x\in \mathcal{X}}\left[\Bar{F}_{k}^{(s)}(x,\hat{\theta}_k) - \Bar{F}_{l}^{(s)}(x,\hat{\theta}_l) \right],
\end{equation*} where $\hat{\theta}_k$ is a consistent estimator of $\theta_{k,0}$ for $k=1,2,...,K$.  The pointwise asymptotic mull distribution of $LMW_{N}$ is a functional of a mean zero Gaussian process when $d_s = 0$ (the ``boundary" case), and degenerates to $-\infty$ when $d_s <0$ (the ``interior" case) under some regularity conditions. For illustration, let $K=2$ and there be no parameter to be estimated.\footnote{The function \textsf{pysdtest.test\_sd()} in the PySDTest package with the `subsampling' argument implements the routine of the LMW test for a two-sample and no parameter case. However, users can implement the test with multiple prospects and parameters by using the additional features provided by PySDTest.} Then, the test statistic $LMW_N$ can be denoted as:
\begin{equation*}
    LMW_N = \sqrt{N} \min \{\sup_{x \in \mathcal{X}} \Bar{D}_{1,2}^{(s)}(x), \sup_{x \in \mathcal{X}}\Bar{D}_{2,1}^{(s)}(x) \}.
\end{equation*} To describe the asymptotic null distribution, define the \textit{contact set} to be:
\begin{equation} \label{eq:contact-set}
    \mathcal{C}_{0} := \{x \in \mathcal{X} : D_{1,2}^{(s)}(x) = 0 \}
\end{equation}
Then, under the null hypothesis $H_0$, 
\begin{align}
    LMW_N \Rightarrow \begin{cases} \min \{\sup_{x \in \mathcal{C}_0} \nu_{1,2}^{(s)}(x), \sup_{x \in \mathcal{C}_0}\nu_{2,1}^{(s)}(x) \} &\text{ if }d_s = 0 \\ - \infty &\text{ if }d_s < 0, \end{cases}
\end{align} where $\nu_{k,l}^{(s)}(\cdot)$ is a mean zero Gaussian process for $k,l = 1,2$ and $k \neq l$. \footnote{This result is based the pointwise asymptotics and hence does not fully describe the behavior of the test statistic in finite samples. See Section 2.2.4 for a result based on the uniform asymptotics.} Under the alternative hypothesis $H_1$ ($d_s > 0)$, $LMW_N \rightarrow_{p} \infty$ where $\rightarrow_p$ denotes convergence in probability. Note that the asymptotic null distribution depends on the true distribution hence is not pivotal.

LMW suggest a \textit{subsampling} procedure to compute the critical values.\footnote{The subsampling method has been proposed by \citet{politis1994stationary}, see \citet{politis1999subsampling} for a comprehensive review.} This method not only circumvents the non-pivotal issue but also yields consistent critical values under a general sampling scheme. Let $b$ be the size of subsamples, satisfyng the condition $b \rightarrow \infty$ and $b/N \rightarrow 0$ as $N \rightarrow \infty$. Let $\alpha \in [0,1]$ be the nominal significance level. Then the subsampling procedure can be conducted as the following steps:
\begin{enumerate}
    \item Compute $LMW_N$ using the full original sample $\mathcal{S}_{N} = \{W_1,...,W_N\}$ where $W_i = \{(Y_{k,i}, Z_{k,i}): k=1,2,...,K \}$.
    \item Generate subsamples $\mathcal{S}_{i,b} = \{W_i,...,W_{i+b-1}\}$ for $i=1,2,...,N-b+1$.
    \item Compute $\{LMW_{i,b}: i=1,2,...,N-b+1\}$ by using the subsamples.
    \item Compute the critical value $g_{N,b}(1-\alpha)$ which is defined to be $(1-\alpha)$ quantiles of the distribution $\{LMW_{i,b}: i=1,2,...,N-b+1\}$.
    \item Reject $H_0$ if $LMW_N > g_{N,b}(1-\alpha)$.
\end{enumerate}
Under Assumptions 1-4 of LMW, the subsampling test is asymptotically valid in the sense
\begin{equation} \label{eq:asymp-valid-LMW}
    \lim_{N \rightarrow \infty}P\left[LMW_N \geq g_{N,b}(1-\alpha) \right] = \begin{cases} \alpha &\text{ if } d_s = 0 \\
     0 &\text{ if } d_s < 0 \\
     1 &\text{ if } d_s > 0. \end{cases}
\end{equation} 

In practice, we need to pick the subsample size $b$ which might affect the test result. Given that there is no compelling theory on the optimal choice of $b$ in the literature, LMW propose three data dependent methods for choosing $b$: \textit{minimum volatility method} \citep{politis1999subsampling}, \textit{mean critical value} and \textit{median critical value}. Let $B_N$ be a set of subsampling size candidates. Then, the minimum volatility method suggests practitioners select one in $B_N$ that minimizes local variations of critical values. The mean and median critical value methods recommend to pick the mean and median of critical values, respectively. However, LMW also suggest computing and drawing a plot of p-values, which checks whether p-values are insensitive to the choice of $b$. This is because, if so, inferences would be robust regardless of the specific value of $b$.

\subsubsection[Contact Set Approach]{Contact Set Approach}

The classical approach imposing the LFC restriction to obtain critical values can be too conservative in practice. This is because it tries to estimate the upper bound of the asymptotic null distributions. To see this,  one can establish that under the null hypothesis, 
\begin{equation} \label{eq:limit-BD}
    BD_N \Rightarrow \sup_{x \in \mathcal{C}_0}\nu_s(x),
\end{equation}
where $\mathcal{C}_0$ is the \textit{contact set} defined in \eqref{eq:contact-set}. Clearly, $\sup_{x \in \mathcal{C}_0}\nu_s(x)$ is stochastically dominated by $\sup_{x \in \mathcal{X}}\nu_s(x)$ because $\mathcal{C}_0 \subseteq \mathcal{X}$. Therefore, the asymptotic size of LFC-based tests achieves the nominal level only when prospects of comparison have equal distributions. This leaves the room for power enhancement.

LSW instead propose the \textit{contact set approach} which estimates the contact set, $\mathcal{C}_0$, and directly mimics the asymptotic null distribution \eqref{eq:limit-BD} for inference. This leads to enhanced power property, compared to the LFC-based procedure. 

LSW consider an integral-type test for arbitrary SD orders with the hypotheses of interests given by \eqref{eq:h0}. In addition to these features, LSW allow unbounded support and prospects depending on finite and infinite dimensional parameters and establish uniform size validity. Their tests can be applied to serially independent observations yet allow cross-sectional dependence between prospects.

Consider the one-sided Cram\'er-von Mises type functional:
\begin{equation*}
    d_s := \int_{\mathcal{X}}\max\{q(x)D_{1,2}^{(s)}(x),0\}^{2}dx,
\end{equation*} where $q(\cdot)$ is a bounded weight function and $D_{1,2}^{(s)}(\cdot)$ is defined as in \eqref{eq:D-kl}.\footnote{LSW also allow practitioners to use a weight function for empirical analysis focusing on a specific part of distributions. For example, one may be interested in comparing income levels below the median of distributions. In this case, the functional would be defined to be
\begin{equation*}
    d_s := \int_{\mathcal{X}}\max\{q(x)D_{1,2}^{(s)}(x),0\}^{2}w(x)dx
\end{equation*} where $w(x)$ is a nonnegative integrable weight function.} The weight function here is to accommodate unbounded supports, which may entail the integrability issue of $F_k^{(s)}$ for $s \geq 2$. If $\mathcal{X}$ is bounded or $s=1$, $q(x) = 1$ is allowed. Otherwise, LSW suggest 
\begin{equation*}
    q(x) = \begin{cases} 1 &\text{ if } x \in [z_1,z_2] \\ 
    a/(a + |x - z_2|^{(s-1)\vee (1+\delta)}) &\text{ if } x > z_2 \\
    a/(a + |x - z_1|^{(s-1)\vee (1+\delta)}) &\text{ if } x < z_1
    \end{cases}
\end{equation*} for $z_1 < z_2$ and $a,\delta > 0$ where $a\vee b$ denotes $\max\{a,b\}$. Then the hypotheses of interests \eqref{eq:h0} can be represented as 
\begin{equation*}
    H_0 : d_s = 0 \text{ v.s. } H_1 : d_s > 0.
\end{equation*} 

Non/semi-parametric models are allowed with the following specification of prospects:
\begin{equation*}
    X_{k}(\theta,\tau) = \varphi_{k}(W;\theta,\tau) \text{ for }  k=1,2
\end{equation*} where $W = (Y,Z) \in \mathbb{R}^{d_W}$ is a random vector and $\varphi(\cdot;\theta,\tau)$ is a real-valued function with parameters $(\theta,\tau)\in \Theta \times \mathcal{T}$. $\Theta$ is a finite-dimensional Euclidean space and $\mathcal{T}$ is an infinite-dimensional space. For example, $X_k := X_k(\theta_0, \tau_0)$ may be specified as the residual from the partially linear regression, i.e. $X_k(\theta,\tau) = Y_k - Z_{k,1}^\intercal \theta - \tau(Z_{k,2})$. Also, we may consider the single-index model such as $X_{k} = Y_k - \tau(Z_{k,1}^{\intercal}\theta)$. 

Let $X_{k,i}(\theta,\tau)  = \varphi_{k}(W_i;\theta,\tau)$ and $\{W_i : i=1,2,...,N\}$ be a random sample. Define the integrated empirical CDF as
\begin{equation*}
    \Bar{F}_{k}^{(s)}(x,\theta,\tau) := \frac{1}{N(s-1)!}\sum_{i=1}^{N}\left[x - X_{k,i}(\theta,\tau)\right]_{+}^{s-1} \text{ for } k=1,2
\end{equation*} and their differences as
\begin{equation*}
    \Bar{D}_{1,2}^{(s)}(x,\theta,\tau) := \Bar{F}_{1}^{(s)}(x,\theta,\tau) - \Bar{F}_{2}^{(s)}(x,\theta,\tau).
\end{equation*} LSW propose the integral type test statistic
\begin{equation}  \label{eq:lsw-test-stat}
    LSW_{N} = N\int_{\mathcal{X}}\max\big\{ q(x)\Bar{D}_{1,2}^{(s)}(x,\hat{\theta},\hat{\tau}),0\big\}^{2}dx,
\end{equation} where $(\hat{\theta},\hat{\tau})$ is a consistent estimator of $(\theta,\tau)$. 

For brevity, we describe the testing procedure under the case when there are no estimated parameters, $q(x)=1$, and $\mathcal{X}$ is bounded. Thus, we consider the following simplified test statistic
\begin{equation*}
    LSW_{N} = N\int_{\mathcal{X}}\max\big\{ \Bar{D}_{1,2}^{(s)}(x),0\big\}^{2}dx
\end{equation*} where $\Bar{D}_{1,2}^{(s)}(\cdot)$ and $\Bar{F}_{k}^{(s)}(\cdot)$ are defined to be the same as in \eqref{eq:bar-D-kl} and \eqref{eq:empirical-opreator}, respectively. Denote the set of probabilities that satisfy the null hypothesis as $\mathcal{P}_0$. Also, let $\mathcal{P}_{00} := \{P \in \mathcal{P}_0 : \lambda(C_0) > 0 \}$ where $\lambda$ is the Lebesgue measure. Put it differently, $\mathcal{P}_{00}$ is the subset of $\mathcal{P}_0$ with positive Lebesgue measure on the contact set. 

We shall describe the asymptotic behavior of $LSW_N$ for a given $P \in \mathcal{P}_0$. With some regularity conditions (Assumptions 1-3 of LSW),
\begin{align*}
    LSW_N &= \int_{\mathcal{X}}\max\big\{\sqrt{N}\left[\Bar{D}_{1,2}^{(s)}(x) - D_{1,2}^{(s)}(x) \right] + \sqrt{N}D_{1,2}^{(s)}(x),0\big\}^{2}dx \\ 
    &= \int_{\mathcal{C}_0}\max\big\{\sqrt{N}\left[\Bar{D}_{1,2}^{(s)}(x) - D_{1,2}^{(s)}(x) \right], 0\big\}^{2}dx \\ 
    &+ \int_{\mathcal{X}\setminus\mathcal{C}_0}\max\big\{\sqrt{N}\left[\Bar{D}_{1,2}^{(s)}(x) - D_{1,2}^{(s)}(x) \right] + \sqrt{N}D_{1,2}^{(s)}(x),0\big\}^{2}dx \\
    &= \int_{\mathcal{C}_0}\max\big\{\sqrt{N}\left[\Bar{D}_{1,2}^{(s)}(x) - D_{1,2}^{(s)}(x) \right], 0\big\}^{2}dx + o_{p}(1) \\
    &\Rightarrow \int_{\mathcal{C}_0}\max\{\nu_{s}(x),0\}^{2}dx,
\end{align*} where $\nu_{s}(\cdot)$ is a mean zero Gaussian process. The second equality comes from the definition of the contact set. The third equality holds because $\sqrt{N}D_{1,2}^{(s)}(x) \rightarrow -\infty$ on $\mathcal{X}\setminus \mathcal{C}_0$ under the null hypothesis. Lastly, the weak convergence follows from $\sqrt{N}\left[\Bar{D}_{1,2}^{(s)}(\cdot) - D_{1,2}^{(s)}(\cdot)\right] \Rightarrow \nu_{s}(\cdot)$ and the continuous mapping theorem. On the other hand, for $P \in \mathcal{P}_0 \setminus \mathcal{P}_{00}$, $\lambda(\mathcal{C}_0) = 0$ gives us $LSW_N \Rightarrow 0$. Therefore, we can summarize the asymptotic results under the null hypothesis as
\begin{equation*}
    LSW_N \Rightarrow \begin{cases} \int_{\mathcal{C}_0}\max\{\nu_{s}(x),0\}^{2}dx &\text{ if } P \in \mathcal{P}_{00} \\ 
    0 &\text{ if } P \in \mathcal{P}_{0}\setminus\mathcal{P}_{00} \end{cases}
\end{equation*} LSW call $\mathcal{P}_{00}$ as the set of \textit{boundary} points and $\mathcal{P}_{0} \setminus \mathcal{P}_{00}$ as the set of \textit{interior} points. The notable thing is that there is a discontinuity in the asymptotic null distributions, depending on the true probability. Since there is no discontinuity in finite samples, the above discontinuity indicates that the \textit{uniform} asymptotics, rather than \textit{pointwise} asymptotics, would provide better approximation and complete asymptotic behaviors.  

Let $\alpha$ be the nominal significance level and $E_{P}$ be the expectation under the probability $P$. Then, the definitions are
\begin{definition}
(a) A test $\rho_{\alpha}$ has an asymptotically (uniformly) valid size if
\begin{equation} \label{eq:asymp-unif-valid}
    \limsup_{N \rightarrow \infty}\sup_{P \in \mathcal{P}_{0}}E_{P}\rho_{\alpha} \leq \alpha
\end{equation}
(b)  A test $\rho_{\alpha}$ has an asymptotically exact size if it satisfies \eqref{eq:asymp-unif-valid} and there exists a non-empty set $\mathcal{P}_{0}' \subset \mathcal{P}_{0}$ such that
\begin{equation} \label{eq:asymp-exact}
    \limsup_{N \rightarrow \infty}\sup_{P \in \mathcal{P}_{0}'}|E_{P}\rho_{\alpha} - \alpha| = 0
\end{equation}
(c) A test $\rho_{\alpha}$ is called asymptotically similar on $\mathcal{P}_{0}'$ if it satisfies \eqref{eq:asymp-exact}.
\end{definition}

The notable thing is that LFC-based testing procedures might be asymptotically non-similar hence biased in large samples against a large set of alternatives. The bootstrap procedure proposed by LSW is asymptotically similar on a larger subset of $\mathcal{P}_0$ in which the LFC is a special case. Therefore, the procedure is asymptotically biased against a smaller class of alternatives than LFC-based tests and hence is more powerful. 

Let $B$ denote the number of bootstrap repetitions and $\alpha$ be the nominal significance level. The procedure can be described as follows:
\begin{enumerate}
    \item Compute $LSW_N$ using the original sample $\{W_1,...,W_N\}$, where $W_i = \{(Y_{k,i}, Z_{k,i}): k=1,2\}$.
    \item Draw $\{W_{i,b}^{*} : i = 1,...,N\}$ with replacement from $\{W_{i}:i=1,...,N \}$.
    \item Compute estimates $\theta_{b}^*$ and $\tau_{b}^*$ using $\{W_{i,b}^{*} : i = 1,...,N\}$.
    \item Compute $X_{k,i,b}^* = \varphi_{k}(W_{i,b}^*;\theta_{b}^*, \tau_{b}^*)$ for $k=1,2$.
    \item Compute 
    \begin{align*}
        \Bar{D}_{1,2,b}^{(s),*}(x) &= \frac{1}{N}\sum_{i=1}^{N}\left[h_{x}(X_{1,i,b}^*) - h_{x}(X_{2,i,b}^*) -  \frac{1}{N}\sum_{i=1}^{N}\{h_{x}(\hat{X}_{1,i}) - h_{x}(\hat{X}_{2,i})\}\right], \\
         \text{ where }& \hat{X}_{k,i} = \varphi_{k}(W_{i};\hat{\theta}, \hat{\tau}) \text{ for } k=1,2 \text{ and } h_{x}(z) := \frac{(x-z)^{(s-1)}1\{z \leq x\}q(x)}{(s-1)!}.
    \end{align*}
    \item Take a sequence $c_N$ such that $c_N \rightarrow 0$ and $c_N \sqrt{N} \rightarrow \infty$ and estimate the contact set as
    \begin{equation*} \label{eq:contact-set-estimation}
        \hat{\mathcal{C}} = \{x \in \mathcal{X} : q(x)|\Bar{D}_{1,2}(x)| < c_N\}.
    \end{equation*}
    \item Compute 
    \begin{equation*}
        LSW_{N,b}^* = 
        \begin{cases}
        \int_{\hat{\mathcal{C}}}\max\{q(x)\sqrt{N}\Bar{D}_{1,2,b}^{(s),*}(x),0\}^{2}dx, &\text{ if } \lambda(\hat{\mathcal{C}}) > 0\\
        \int_{\mathcal{X}}\max\{q(x)\sqrt{N}\Bar{D}_{1,2,b}^{(s),*}(x),0\}^{2}dx, &\text{ if } \lambda(\hat{\mathcal{C}}) = 0.
        \end{cases}
    \end{equation*}
    \item Repeat step 2-7 for $b=1,2,...,B$ and compute the critical value \begin{equation*}
        c_{\alpha,N,B}^* := \inf \{ B^{-1}\sum_{b=1}^{B}1\{LSW_{N,b}^* \leq t\} \geq 1-\alpha \}.
    \end{equation*}
\end{enumerate}
Note that $W_i = (X_{1,i},X_{2,i})^{\intercal}$ and step 3-4 should be omitted if there are no estimated parameters. 

We shall introduce some asymptotic properties of the contact set approach. Suppose that Assumptions 1-4 of LSW hold. Under $H_0$, 
\begin{equation} \label{eq:contact-size}
    \limsup_{N \rightarrow \infty}\sup_{P \in \Tilde{\mathcal{P}}_0} P (LSW_N > c_{\alpha,N,\infty}^*) \leq \alpha
\end{equation} for some $\Tilde{\mathcal{P}}_0 \subset \mathcal{P}_0$ where $c_{\alpha,N,\infty}^*$ is the critical value from the ideal bootstrap ($B = \infty$). In addition, for some $\Tilde{\mathcal{P}}_{00,N} \subset \Tilde{\mathcal{P}}_0$,
\begin{equation} \label{eq:contact-exact}
    \limsup_{N \rightarrow \infty}\sup_{P \in \Tilde{\mathcal{P}}_{00,N}}| P (LSW_N > c_{\alpha,N,\infty}^*) - \alpha| = 0.
\end{equation} Equation \eqref{eq:contact-size} shows that the test has asymptotically vaild size uniformly over a class of ``regular'' probabilities $\Tilde{\mathcal{P}}_{0} \subset \mathcal{P}_0$. In addition, equation \eqref{eq:contact-exact} indicates that test has asymptotically exact size on $\Tilde{\mathcal{P}}_{00,N} \subset \Tilde{\mathcal{P}}_0$. On the other hand, the test is consistent against a fixed alternative. That is, for fixed alternative $P \notin \mathcal{P}_0$ such that $\int_{\mathcal{X}} \max\{q(x)D_{1,2}(x),0 \}^{2}dx > 0$, 
\begin{equation*} \label{eq:contact-consist}
    \lim_{N \rightarrow \infty} P(LSW_N >  c_{\alpha,N,\infty}^*) = 1
\end{equation*}
In addition, Theorem 4 of LSW shows that the contact set approach is locally more powerful than the LFC-based tests.

\subsubsection[Selective Recentering Approach]{Selective Recentering Approach}

DH propose the \textit{selective recentering approach} to enhance the power performance of BD tests. The main idea is similar to the contact set approach: DH try to directly approximate the limit distribution of $BD_N$, $\sup_{x \in \mathcal{C}_0} \nu_s(x)$ defined in \eqref{eq:limit-BD}, to attain the critical values. For this end, DH extend the recentering method introduced by \citet{hansen2005test}. It is also closely related to the generalized moment selection approach of \citet{andrews2013inference} in the moment inequality literature.

 DH  also consider the hypotheses of interest given by \eqref{eq:h0} and they consider the Barrett-Donald test statisitc $BD_N$. They consider a sampling scheme similar to that of BD, but further discuss extension of their method to deal with weakly dependent data and mutual dependence between $X_1$ and $X_2$.

Let us briefly describe the selective recentering approach. Let
\begin{equation} \label{eq:recenter-pop}
    \mu(x) :=\min\{D_{1,2}^{(s)}(x),0\}.
\end{equation} Then, we have that the asymptotic null distribution of
\begin{equation} \label{eq:BD0}
    BD^{0}_{N} = 
    \sup_{x\in \mathcal{X}}\big[\sqrt{N}\big( \Bar{D}_{1,2}^{(s)}(x) - D_{1,2}^{(s)}(x) \big) + \sqrt{N}\mu(x)\big]
\end{equation} is the same as that of $BD_N$ because $\mu(x) = D_{1,2}^{(s)}(x)$ under the null. DH propose a resampling procedure that tries to mimic \eqref{eq:BD0} using the sample analogue of \eqref{eq:recenter-pop}. That is, first take a sequence $a_N$ of negative numbers such that $a_N \rightarrow -\infty$ and $a_N/\sqrt{N} \rightarrow 0$ and approximate \eqref{eq:recenter-pop} via the \textit{recentering function}:
\begin{equation*} \label{eq:recentering-function}
    \hat{\mu}_{N}(x) :=     \Bar{D}^{(s)}_{1,2} \cdot 1 \left\{ \Bar{D}^{(s)}_{1,2} < a_N/\sqrt{N} \right\}
\end{equation*} 
Under regularity conditions, it is a uniformly consistent estimator in the sense that
\begin{equation*}
    \sup_{x \in \mathcal{X}}|\hat{\mu}_{N}(x) - \mu(x) | \rightarrow_p 0
\end{equation*}

Write the simulated processes defined in \eqref{eq:mp-simul-BD}, \eqref{eq:BD-pool}, and \eqref{eq:BD-recenter} as
\begin{align*}
    \nu_{s,MP}^{*} &= \mathcal{J}_s(x, \mathcal{B}_{1}^* \circ \Bar{F}_1) - \mathcal{J}_s(x, \mathcal{B}_{2}^* \circ \Bar{F}_2),  \\
    \nu_{s,pool}^{*} &= \mathcal{J}_s(x, \Bar{F}^*_1) - \mathcal{J}_s(x, \Bar{F}^*_2),  \\
    \nu_{s,recenter}^{*} &= (\mathcal{J}_s(x, \Bar{F}^*_1) - \mathcal{J}_s(x, \Bar{F}_1)) - ( \mathcal{J}_s(x, \Bar{F}^*_2) - \mathcal{J}_s(x, \Bar{F}_2)). 
\end{align*} 
To compute the critical values, DH suggest using the bootstrap distributions 
\begin{equation} \label{eq:sr4} 
    DH_{N,k}^{*} = \sup_{x \in \mathcal{X}}\{\nu_{s,k}^{*}(x) + \sqrt{N}\hat{\mu}_{N}(x)\} \text{ for } k=\{MP, pool, recenter\}.
\end{equation} Letting $\hat{c}_{\alpha,k}$ be the $(1-\alpha)$ quantile of $DH_{N,k}^{*}$, DH suggest the following critical value: 
\begin{equation*}
    \hat{c}_{\alpha,\eta,k} := \max\{\hat{c}_{\alpha,k},\eta\},
\end{equation*} where $\eta$ is a small positive number such as $10^{-6}$. $\eta$ is introduced here to accommodate the situation where both the original test statistic and the bootstrap test statistic degenerate to zero, such as the interior case $D_{1,2}^{(2)} (x) < 0$ for all $x \in \mathcal{X}$.  

In practice, two tuning parameters, $a_N$ and $\eta$, need to be selected. Based on the simulation experiments conducted by DH, the specific choice of $\eta$ is not very important as long as $\eta$ is a small positive number. However, $a_N$ should be chosen carefully because the choice of $a_N$ may affect the finite sample performance of the test. DH suggest using $a_N = -0.1\sqrt{\log\log(N)}$ which performed well in their simulations and empirical studies.

As regard to the asymptotic properties, DH show that under the null hypothesis,
\begin{align*}
    \lim_{N \rightarrow \infty}P(BD_N > \hat{c}_{\alpha,\eta,k}) = 0 \text{ if } D_{1,2}^{(s)}(x) < 0 \text{ for all } x \in \mathcal{X} \\
    \lim_{N \rightarrow \infty}P(BD_N > \hat{c}_{\alpha,\eta,k}) = \alpha \text{ if } D_{1,2}^{(s)}(x) = 0 \text{ for all } x \in \mathcal{X}
\end{align*} For a fixed alternative, the test is shown to be consistent:
\begin{equation*}
    \lim_{N \rightarrow \infty}P(BD_N > \hat{c}_{\alpha,\eta,k}) = 1.
\end{equation*} 

\subsubsection[Numerical Delta Method]{Numerical Delta Method}

As an alternative method for attaining critical values, we can apply an inference method applicable to directionally differentiable functions. \citet{fang2019inference} show the asymptotic validity of an inference method that employs a consistent estimate of the first order directional derivative when the target parameters are Hadamard directionally differentiable. Complementing their idea, \citet{hong2018numerical} propose the method using numerical derivatives for approximating the limit distributions which eases the burden of analytical calculation.\footnote{Both \citet{fang2019inference} and \citet{hong2018numerical} are extensions of the pioneering work of \citet{dumbgen1993nondifferentiable} that studies a validity of the bootstrap under a lack of differentiability.} Even though \citet{fang2019inference} discuss the testing for SD as an application of their methods, they do not provide a formal result let alone the software for the SD tests.

We briefly discuss the NDM, which also has enhanced power properties compared to the classical LFC-based tests. The power enhancement is possible because the NDM also directly approximates the asymptotic null distribution, as in the contact set approach and selective recentering approach. Theoretical validity of applying the NDM to SD tests is discussed by \citet{lee2023testing}. To be specific, \citet{lee2023testing} introduce the test for time stochastic dominance (time SD) with application of the NDM and the static SD concept is regarded as a special case of the time SD concept with a single time period.  

To describe testing for SD by the NDM, we first introduce the definition of \textit{Hadamard directionally differentiability} of a map between Banach spaces \citep{shapiro1990concepts}.
\begin{definition}
Let $\mathbb{D}$ and $\mathbb{E}$ be Banach spaces and $\phi:\mathbb{D}_{\phi} \subset \mathbb{D} \rightarrow \mathbb{E}$. The map $\phi$ is said to be Hadamard directionally differentiable at $\theta \in \mathbb{D}_{\phi}$ tangentially to a set $\mathbb{D}_{0} \subset \mathbb{D}$ if there exists a continuous map $\phi'_{\theta} : \mathbb{D}_0 \rightarrow \mathbb{E}$ such that
\begin{equation*}
    \lim_{N \rightarrow \infty} \left\Vert \frac{\phi(\theta + t_N h_N) - \phi(\theta)}{t_N} - \phi_{\theta}'(h) \right\Vert_{\mathbb{E}} = 0
\end{equation*} for all sequences $\{h_N\} \subset \mathbb{D}$ and $\{t_N\} \subset \mathbb{R}_{+}$ such that $t_N \downarrow 0$, $h_N \rightarrow h \in \mathbb{D}_0$ as $N \rightarrow \infty$ and $\theta + t_N h_N \in \mathbb{D}_{\phi}$ for all $N$. 
\end{definition} We say $\phi'_{\theta}(h)$ is (first-order) Hadamard directional derivative at $\theta$ in direction $h$. In addition, if $\phi'_{\theta}$ is linear, then $\phi$ is said to be Hadamard differentiable at $\theta \in \mathbb{D}_{\phi}$ tangentially to a set $\mathbb{D}_{0} \subset \mathbb{D}$.

Theorem 2.1 of \citet{fang2019inference} shows that if $r_N(\hat{\theta}_N - \theta_0) \Rightarrow \mathbb{G}_0$ in $\mathbb{D}$ for some $r_N \rightarrow \infty$ where $\mathbb{G}_0$ is tight, then $r_N\left[\phi(\hat{\theta}_N) - \phi(\theta_0)\right] \Rightarrow \phi'_{\theta_{0}}(\mathbb{G}_0)$. On top of this result, Theorem 3.2 of \citet{fang2019inference} shows that $\phi'_{\theta_{0}}(\mathbb{G}_0)$ can be approximated by $\phi'_{N}(\mathbb{Z}_N^*)$ where $\mathbb{Z}_N^*$ and $\phi_{N}'$ are consistent estimators of $\mathbb{G}_0$ and $\phi'_{\theta_{0}}$, respectively. \citet{hong2018numerical} suggest the NDM which numerically approximates $\phi'_{\theta_{0}}(\mathbb{G}_0)$ hence does not need analytical calculation of $\phi'_{\theta_0}$. \citet{hong2018numerical} suggest the estimation of $\phi'_{\theta_0}$ by using the following numerical derivatives
\begin{equation} \label{eq:approx-ndm-1}
    \Tilde{\phi}'_{N}(\mathbb{Z}_N^*) := \frac{\phi(\hat{\theta}_N + \epsilon_N \mathbb{Z}_N^*) - \phi(\hat{\theta}_N)}{\epsilon_N}
\end{equation} where $\epsilon_N$ is a step size satisfying $\epsilon_N \rightarrow 0$ and $r_N\epsilon_N \rightarrow \infty$ as $N \rightarrow \infty$.

The NDM can be applied to the SD testing problem. Consider the testing with the $L_1$-type test statistic and our hypothesis of interest is the same as \eqref{eq:h0}. First, denote 
\begin{equation*}
  l^{\infty}(\mathcal{X)} := \{f : \mathcal{X} \rightarrow \mathbb{R} \text{ such that } \left\Vert f \right\Vert_{\infty} < \infty \} \text{ and } \left\Vert f \right\Vert_{\infty} = \sup_{x \in \mathcal{X}} f(x).
\end{equation*}
For any $\theta \in l^{\infty}(\mathcal{X}) \times l^{\infty}(\mathcal{X})$, $\phi : l^{\infty}(\mathcal{X}) \times l^{\infty}(\mathcal{X}) \rightarrow \mathbb{R}$ is defined as
\begin{equation*}
    \phi(\theta) := \int_{\mathcal{X}} \max\{\theta_{1}(x) - \theta_{2}(x),0\}dx.
\end{equation*}  We then may write 
\begin{equation*}
    \phi(\theta_0) = \int_{\mathcal{X}} \max\{F_{1}^{(s)}(x) - F_{2}^{(s)}(x),0\}dx,
\end{equation*} where $ \theta_0 = (F_{1}^{(s)}(x), F_{2}^{(s)}(x)) \in l^{\infty}(\mathcal{X}) \times l^{\infty}(\mathcal{X})$. It can be seen that $\phi(\cdot)$ is Hadamard directionally differentiable with its derivative given as
\begin{equation*}
    \phi'_{\theta}(h) = \int_{\mathcal{C}_{+}(\theta)}(h_{1}(x) - h_{2}(x))dx + \int_{\mathcal{C}_{0}(\theta)}\max\{h_{1}(x) - h_{2}(x),0\}dx,
\end{equation*} where
\begin{align*}
    \mathcal{C}_{+}(\theta) &:= \{x \in \mathcal{X} : \theta_1(x) > \theta_2(x) \} \\
    \mathcal{C}_{0}(\theta) &:= \{x \in \mathcal{X} : \theta_1(x) = \theta_2(x) \}.
\end{align*} Note that $\mathcal{C}_{0}(\theta_0)$ is equivalent to the contact set $\mathcal{C}_0$ in \eqref{eq:contact-set}. In addition, we have $\mathcal{C}_{+}(\theta_0) = \emptyset$ under the null hypothesis. 

Letting $\hat{\theta}_N = (\Bar{F}_{1}^{(s)}(\cdot),\Bar{F}_{2}^{(s)}(\cdot))$, we may write the $L_1$-type test statistic as $r_N\phi(\hat{\theta}_N)$, where $r_N =\sqrt{N}$. In addition, we know that $r_N(\Bar{F}_{k}^{(s)}(\cdot) - F_{k}^{(s)}(\cdot))$ weakly converges to a tight Gaussian process. If we let $\mathbb{Z}_N^* = r_N(\Bar{F}_{k}^{(s),*}(\cdot) - \Bar{F}_{k}^{(s)}(\cdot))$, where $\Bar{F}_{k}^{(s),*}$ is the bootstrap sample analogue of $\Bar{F}_{k}^{(s)}$, then we can approximate the limit distribution of $r_N\phi(\hat{\theta}_N)$ by employing \eqref{eq:approx-ndm-1}.

There are some cases in which the first-order derivatives may degenerate, such as the the $L_2$-type test statistic. In this case, we can employ the second-order NDM \citep{hong2018numerical}. $\phi_{\theta_0}''(h)$ is said to be the second order Hadamard directional derivative at $\theta_0 \in \mathbb{D}_{\phi}$ in the direction $h$ if
\begin{equation*}
    \lim_{N \rightarrow \infty}\left\Vert \frac{\phi(\theta_0 + t_N h_N) - \phi(\theta_0) - t_N \phi'_{\theta_0}(h_N)}{\frac{1}{2}t_N^2} - \phi_{\theta_0}''(h) \right\Vert_{\mathbb{E}} = 0
\end{equation*} for all sequences $\{h_N\} \subset \mathbb{D}$ and $\{t_N\} \subset \mathbb{R}_{+}$ such that $t_N \downarrow 0$, $h_N \rightarrow h \in \mathbb{D}_0$ as $N \rightarrow \infty$ and $\theta + t_N h_N \in \mathbb{D}_{\phi}$ for all $N$. Theorem 4.1 of \citet{hong2018numerical} shows that if $r_N(\hat{\theta}_n - \theta_0) \Rightarrow \mathbb{G}_0$ for some tight process $\mathbb{G}_0$, then 
\begin{equation*}
    r_N^2\left[\phi(\hat{\theta}_N) - \phi(\theta_0) - \phi_{\theta_0}'(\hat{\theta}_N - \theta_0)  \right] \Rightarrow \frac{1}{2}\phi''_{\theta_0}(\mathbb{G}_0)
\end{equation*} When the first-order derivatives degenerate, $\phi'_{\theta_0}(h) = 0$, it is possible to approximate $\frac{1}{2}\phi''_{\theta_0}(\mathbb{G}_0)$ by 
\begin{equation} \label{eq:approx-ndm-2}
    \frac{1}{2}\Tilde{\phi}_{\theta_0}''(\mathbb{Z}_N^*) := 
    \begin{cases} 
    \frac{\phi(\hat{\theta}_N + \epsilon_N\mathbb{Z}_N^*)- \phi(\hat{\theta}_N) }{\epsilon^2_N} \text{ or } \\
    \frac{\phi(\hat{\theta}_N + 2\epsilon_N\mathbb{Z}_N^*)- 2\phi(\hat{\theta}_N - \epsilon_N\mathbb{Z}_N^*) + \phi(\hat{\theta}_N) }{2\epsilon^2_N} 
    \end{cases}
\end{equation} where $\epsilon_N$ satisfies $\epsilon_N \rightarrow 0$ and $r_N\epsilon_N \rightarrow \infty$ as $N \rightarrow \infty$.
In the case of the $L_{2}$-type test statistic, which is the simplified version of $LSW_N$ in \eqref{eq:lsw-test-stat}, we may take 
\begin{align*}
    \phi(\theta) &:= \int_{\mathcal{X}} \max\{\theta_{1}(x) - \theta_{2}(x),0\}^{2}dx \\
    \phi(\theta_0) &= \int_{\mathcal{X}} \max\{F_{1}^{(s)}(x) - F_{2}^{(s)}(x),0\}^{2}dx
\end{align*} Then we have $\phi_{\theta_0}'(h) = 0$ under the null hypothesis but
\begin{equation*}
    \phi''_{\theta_0}(h) = \int_{ \mathcal{C}_0(\theta_0)}\max\{h_{1}(x) - h_{2}(x),0\}^{2}dx
\end{equation*} holds. Thus, similarly to the first-order NDM case, we can approximate the limit distribution of $r_N^2 \phi(\hat{\theta}_N)$ using \eqref{eq:approx-ndm-2}.

\section[The Package: PySDTest]{The Package: PySDTest} \label{sec:pysdtest}

\subsection{Installation}

It is required to install the package, PySDTest, for both Python and Stata environments. The package is listed on the Python Package Index (PyPI) which is a repository of softwares for the Python programming language. If Python is installed on your computer, a user can install PySDTest by entering:

\begin{lstlisting}
pip install PySDTest  
\end{lstlisting} in Windows cmd or Mac (or Linux) terminal.\footnote{Alternatively, users can access source codes of the package through \href{https://github.com/lee-kyungho/pysdtest}{https://github.com/lee-kyungho/pysdtest}} For details about installing a Python package, see \href{https://packaging.python.org/tutorials/installing-packages/}{Python Package User Guide in PyPI}.

The package is written based on Python numpy module with version $\geq 1.26.4$. Therfore, it is also required to install numpy for employing the package.

To install the Stata module, \texttt{pysdtest}, users can use the following command in Stata:

\begin{lstlisting}
net install pysdtest, from("https://raw.githubusercontent.com/lee-kyungho/pysdtest/main/Stata") replace \end{lstlisting}

\subsection{Implementation in Python}

This section describes how to use the package PySDTest in the Python environment. Notably, Stata with version $\geq$ 16.0 supports running the package within Stata. For more details on using Python in Stata, please refer to the \href{https://www.stata.com/stata16/python-integration/}{Python integration feature} in Stata.

First, users need to import the package through the following code in the Python environment in order to use the package:
\begin{CodeChunk}
\begin{CodeInput}
>>> import pysdtest
\end{CodeInput}
\end{CodeChunk}

PySDTest implements comprehensive SD tests developed in the literature. In addition, users can freely choose procedures suitable to their objects of study. In this section, we describe the details of how to implement each testing procedure.

We first generate simulated data as:
\begin{CodeChunk}
\begin{CodeInput}
>>> import numpy as np
>>> mu1 , mu2 , sigma1 , sigma2 = 0, 0 , 1, 1
>>> n = 500
>>> np.random.seed(0)
>>> sample1 = mu1 + sigma1 * np.random.randn(n)
>>> sample2 = mu2 + sigma2 * np.random.randn(n)
\end{CodeInput}
\end{CodeChunk}

\subsubsection[BD and LMW Tests by \textsf{pysdtest.test\_sd()}]{BD and LMW tests by \textsf{pysdtest.test\_sd()}}

The function \textsf{pysdtest.test\_sd()} enables the user to implement the tests proposed by BD and LMW. The structure of the function \textsf{pysdtest.test\_sd} is given as:

\begin{CodeChunk}
\begin{CodeInput}
pysdtest.test_sd(sample1, sample2, ngrid, s, resampling, b1 = None, 
b2 = None, nboot = 200, quiet = False)
\end{CodeInput}
\end{CodeChunk}

\begin{itemize}
    \item \texttt{sample1} and \texttt{sample2} are 1-dimensional NumPy arrays.
    \item \texttt{ngrid} determines the number of grid points for computing the value on the support.\footnote{Increasing \texttt{ngrid} enhances precision but raises computational costs.}
    \item \texttt{s} argument specifies the stochastic order of the hypothesis.
    \item \texttt{resampling} argument defines the resampling method: 
    \begin{itemize}
        \item \texttt{bootstrap}: Implements the recentered bootstrap method as described in BD.
        \item \texttt{subsampling}: Follows the method suggested by LMW, requiring subsampling sizes \(b_1\) and \(b_2\) for each sample.
        \item \texttt{paired\_bootstrap}: Uses a paired recentered bootstrap \((X_i, Y_i)\) to consider dependencies.
    \end{itemize}
    \item \texttt{nboot}: Sets the number of bootstrapping iterations. The default value is 200 for both \texttt{bootstrap} and \texttt{paired\_bootstrap}
    \item \texttt{quiet}: Specifies whether to print output results or not.
\end{itemize}

The following lines of codes implement the testing procedure by BD:

\begin{CodeChunk}
\begin{CodeInput}
>>> testing_normal =  pysdtest.test_sd(sample1, sample2, ngrid = 100, 
s = 1, resampling = 'bootstrap')
>>> testing_normal.testing()
\end{CodeInput}
\begin{CodeOutput}
#--- Testing for Stochastic Dominance  -----#

* H0 : sample1 first order SD sample2

#-------------------------------------------#

*** Test Setting ***
* Resampling method         = bootstrap
* SD order                  =      1
* # of (sample1)            =    500 
* # of (sample2)            =    500
* # of bootstrapping        =    200
* # of grid points          =    100

#-------------------------------------------#

*** Test Result ***
* Test statistic            = 0.2214
* Significance level        =  0.05
* Critical-value            = 1.2017
* P-value                   = 0.8600
* Time elapsed              =  0.54 Sec

\end{CodeOutput}
\end{CodeChunk}

Users can run the subsampling procedure by giving \texttt{subsampling} as an input for the \texttt{resampling} argument:

\begin{CodeChunk}
\begin{CodeInput}
>>> testing_normal_sub =  pysdtest.test_sd(sample1, sample2, ngrid = 100, 
s = 1, resampling = 'subsampling', b1=40,b2=40)
>>> testing_normal_sub.testing()
\end{CodeInput}
\begin{CodeOutput}
#--- Testing for Stochastic Dominance  -----#

* H0 : sample1 first order SD sample2

#-------------------------------------------#

*** Test Setting ***
* Resampling method         = subsampling
* SD order                  =      1
* # of (sample1)            =    500 
* # of (sample2)            =    500
* # of (subsample1)         =     40
* # of (subsample2)         =     40

#-------------------------------------------#

*** Test Result ***
* Test statistic            = 0.2214
* Significance level        =  0.20
* Critical-value            = 0.8944
* P-value                   = 0.8482
* Time elapsed              =  0.11 Sec
\end{CodeOutput}
\end{CodeChunk}

\subsubsection[Contact Set Approach by \textsf{pysdtest.test\_sd\_contact()}]{Contact Set Approach by \textsf{pysdtest.test\_sd\_contact()}}

The function \textsf{pysdtest.test\_sd\_contact()} in PySDTest implements the contact set approach proposed by LSW.\footnote{To be specific, the case when there is no estimated parameters and $q(x) = 1$ for all $x$ is implemented by \textsf{pysdtest.test\_sd\_contact()}.}  The structure of the function is written as
\begin{CodeChunk}
\begin{CodeInput}
pysdtest.test_sd_contact(sample1, sample2, ngrid, s, resampling, 
nboot=200, c=0.75, quiet = False)
\end{CodeInput}
\end{CodeChunk}

Arguments \texttt{sample1}, \texttt{sample2}, \texttt{ngrid},
 \texttt{s}, \texttt{resampling},  \texttt{b1}, \texttt{b2}, \texttt{nboot}, and \texttt{quiet} are the same as \textsf{pysdtest.test\_sd()}.  

We need to take a sequence $c_N$ for the contact set estimation in the step 6 of LSW. Following the recommendation by LSW, we set
\begin{equation*} \label{eq:c-N-pysd}
    c_N = \frac{c\log(\log(N))}{\sqrt{N}}
\end{equation*}
where $c$ is some positive constant. If two sample sizes are different, $N = (N_1 + N_2)/2$ holds. The argument \texttt{c} designates the constant $c$. We set the default value of $c$ to be $0.75$.

Users can implement the LSW test by:
\begin{CodeChunk}
\begin{CodeInput}
>>> testing_normal_contact =  pysdtest.test_sd_contact(sample1, sample2, 
ngrid = 100, s = 1, resampling = 'bootstrap')
>>> testing_normal_contact.testing()
\end{CodeInput}
\begin{CodeOutput}
#--- Testing for Stochastic Dominance  -----#

* H0 : sample1 first order SD sample2
* Contact Set Approach

#-------------------------------------------#

*** Test Setting ***
* Resampling method        = bootstrap
* SD order                 =      1
* # of (sample1)           =    500 
* # of (sample2)           =    500
* # of bootstrapping       =    200
* # of grid points         =    100

# Tuning parameter -------
* c                        = 0.7500

#-------------------------------------------#

*** Test Result ***
* Test statistic          = 0.1495
* Significance level      =  0.05
* Critical-value          = 18.9272
* P-value                 = 0.8850
* Time elapsed            =  0.52 Sec
\end{CodeOutput}
\end{CodeChunk}

\subsubsection[Selective Recentering Approach by \textsf{pysdtest.test\_sd\_SR()}]{Selective Recentering Approach by \textsf{pysdtest.test\_sd\_SR()}}

The selective recentering approach of DH with $k=recenter$ (Equation (\ref{eq:sr4}))  can be implemented by \textsf{pysdtest.test\_sd\_SR()}. The structure of the function is
\begin{CodeChunk}
\begin{CodeInput}
pysdtest.test_sd_SR(sample1, sample2, ngrid, s, resampling, nboot=200, 
a=0.1, eta=1e-06, quiet = False)
\end{CodeInput}
\end{CodeChunk}

Arguments \texttt{sample1}, \texttt{sample2}, \texttt{ngrid},
 \texttt{s}, \texttt{resampling},  \texttt{b1}, \texttt{b2}, \texttt{nboot}, and \texttt{quiet} are the same as \textsf{pysdtest.test\_sd()}.  
 
The selective recentering approach additionally requires two tuning parameters: $a_N$ and $\eta$. We follow the recommendation by DH and set $a_N$ to be
\begin{equation*}
    a_N = -a \sqrt{\log(\log (N))},
\end{equation*} where $a$ is some positive constant and $N = N_1 + N_2$. Arguments \texttt{a} and \texttt{eta} set the values of $a$ and $\eta$ with the default given by  $a = 0.1$ and $\eta = 10^{-6}$.

\begin{CodeChunk}
\begin{CodeInput}
>>> testing_normal_SR =  pysdtest.test_sd_SR(sample1, sample2, 
ngrid = 100, s = 1, resampling = 'bootstrap')
>>> testing_normal_SR.testing()
\end{CodeInput}
\begin{CodeOutput}
#--- Testing for Stochastic Dominance  -----#

* H0 : sample1 first order SD sample2
* Selective Recentering Approach

#-------------------------------------------#

*** Test Setting ***
* Resampling method       =  bootstrap
* SD order                =      1
* # of (sample1)          =    500 
* # of (sample2)          =    500
* # of bootstrapping      =    200
* # of grid points        =    100

# Tuning paremeters -------------
* a              	 = 0.1000
* eta            	 = 0.000001
#-------------------------------------------#

*** Test Result ***
* Test statistic          = 0.2214
* Significance level      =  0.05
* Critical-value          = 1.0768
* P-value                 = 0.8200
* Time elapsed            =  0.52 Sec
\end{CodeOutput}
\end{CodeChunk}

\subsubsection[Numerical Delta Method by \textsf{pysdtest.test\_sd\_NDM()}]{Numerical Delta Method by \textsf{pysdtest.test\_sd\_NDM()}}

Users can approximate the limit distribution based on the NDM approach using the function \textsf{pysdtest.test\_sd\_NDM()}. The function has the following structure:
\begin{CodeChunk}
\begin{CodeInput}
pysdtest.test_sd_NDM(sample1, sample2, ngrid, s, resampling, nboot = 200,
epsilon = None, form = "L1", quiet = False)
\end{CodeInput}
\end{CodeChunk}

Three different types of functional are available for users: Kolmogorov-Smirnov (supremum), $L_{1}$, and $L_{2}$ type functionals. This can be set by giving input `KS', `L1', and `L2' for the `form' argument, respectively. We apply the second-order NDM for $L_2$ type testing and the limit distribution is approximated by $\frac{\phi(\hat{\theta}_N + \epsilon_N\mathbb{Z}_N^*)- \phi(\hat{\theta}_N) }{\epsilon^2_N}$.

In addition, the NDM requires users to specify the step size ($\epsilon_{N}$) as a tuning parameter. The user can set the step size by the argument `epsilon'. We set $\epsilon = r_{N}^{-1/16}$ as a default value when nothing is specified by the user. The user can implement the NDM for SD tests as
\begin{CodeChunk}
\begin{CodeInput}
>>> testing_normal_NDM_KS =  pysdtest.test_sd_NDM(sample1, sample2, ngrid = 100,
s = 1, resampling = 'bootstrap', form = 'KS')
>>> testing_normal_NDM_KS.testing()
\end{CodeInput}
\begin{CodeOutput}
#--- Testing for Stochastic Dominance  -----#

* H0 : sample1 first order SD sample2
* Numerical Delta Method
* KS Type Test Statistic 

#-------------------------------------------#

*** Test Setting ***
* Resampling method       = bootstrap
* SD order                =      1
* # of (sample1)          =    500 
* # of (sample2)          =    500
* # of bootstrapping      =    200
* # of grid points        =    100

# Tuning paremeter -------------
* epsilon                 = 0.8415

#-------------------------------------------#

*** Test Result ***
* Test statistic          = 3.1939
* Significance level      =  0.05
* Critical-value          = 0.9805
* P-value                 = 0.0000
* Time elapsed            =  0.51 Sec
\end{CodeOutput}
\end{CodeChunk}

Table \ref{table:common-input} contains common arguments for all testing procedures: BD, LMW, LSW, DH, and NDM. In addition, we summarize additional arguments for each estimation in Table \ref{table:add-input}.

\begin{table}[htbp]
        \centering
        \caption{Common Input Arguments for Testing}
        \resizebox{\textwidth}{!}{
        \begin{tabular}{llcc}
        \toprule
        Argument & Description & Data type \\
        \midrule
        \texttt{sample1}       & $X_1$  & (1-dim) numpy array \\
        \texttt{sample2}       & $X_2$  & (1-dim) numpy array \\
        \texttt{s}           & Stochastic dominance order & int  \\
        \texttt{ngrid}         & Number of grid points & int \\
        \texttt{resampling}    & Resampling methods (\texttt{bootstrap}, \texttt{subsampling}, \texttt{paired\_bootstrap}) & str \\
        \texttt{nboot}         & Number of bootstrapping   & int  \\
        \texttt{b1}          & Size of subsampling for sample1 & int \\
        \texttt{b2}          & Size of subsampling for sample2 & int \\
        \texttt{quiet}          & whether to print output results  & boolean \\
\bottomrule
\end{tabular}}\label{table:common-input}
\end{table} 
\begin{table}[htbp]
        \centering
        \caption{Additional Arguments for each Procedure}
        \resizebox{\textwidth}{!}{
        \begin{tabular}{lllc}
        \toprule
        Argument & Description & Method & Data type \\
        \midrule
        \texttt{c} & Tuning parameter for a sequence $c_{N}$ & Contact set approach (LSW) & float\\
        \texttt{a} & Tuning parameter for a sequence $a_{N}$  & Selective recentering (DH) & float \\
        \texttt{eta} & Tuning parameter $\eta$ for attaining a critical value & Selective recentering (DH) & float \\
        \texttt{epsilon} & Step size $\epsilon_{N}$ for numerical approximation & Numerical delta method (NDM) & float \\
        \texttt{form} & Type of functional (KS, L1, and L2) & Numerical delta method (NDM) & str \\
    
\bottomrule
\end{tabular}}\label{table:add-input}
\end{table}

\subsubsection[Additional Features]{Additional Features}

Every implementation of testing through PySDTest saves results as an object hence users can save the test results. For example, if the user runs the following code
\begin{CodeChunk}
\begin{CodeInput}
>>> results = testing_normal.result
\end{CodeInput}
\end{CodeChunk}
then the \emph{`results'} in the above code is the variable in which the Python dictionary of the test result is allocated. The dictionary contains calculated test statistics, resampled test statstics, critical-value and p-value. 

In addition, every function for testing provides a feature for plotting integrated empirical CDFs. This functionality may help users to visually compare testing objects. The plotting is based on the Python Matplotlib package. The method \textsf{plot\_CDF()} has the following structure
\begin{CodeChunk}
\begin{CodeInput}
plot_CDF(save=False, title=None, label1='sample1', label2='sample2', xlabel='x')
\end{CodeInput}
\end{CodeChunk}

In table \ref{input-cdf}, arguments for \textsf{plot\_CDF()} are listed.
\begin{table}[htbp]
        \centering
        \caption{Input Arguments for \textsf{plot\_CDF()}}
        \resizebox{0.75\textwidth}{!}{
        \begin{tabular}{llc}
        \toprule
        Argument & Description & Data type \\
        \midrule
        \texttt{save}      & Whether to save figure (True or False) & boolean \\
        \texttt{title}     & Title of saving image file & str \\
        \texttt{label1}    & Name of sample1 for legend & str  \\
        \texttt{label2}    & Name of sample2 for legend & str \\
        \texttt{xlabel}    & Name of x-axis & str \\
        \bottomrule
\end{tabular}} \label{input-cdf}
\end{table} 

\begin{figure}[htbp]
    \centering
    \caption{(First order) Plotting Result of \emph{plot\_CDF()}}
    \includegraphics[width=0.75\textwidth]{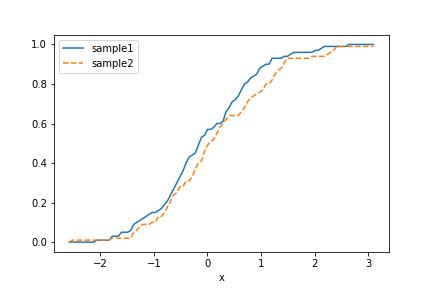}
    \label{fig:plot-cdf}
\end{figure}

For instance, plotting can be conducted by 
\begin{CodeChunk}
\begin{CodeInput}
>>> testing_normal.plot_CDF(save=True, title="Normal distributions.png") 
\end{CodeInput}
\end{CodeChunk}
and we show the corresponding result in Figure \ref{fig:plot-cdf}.

\begin{figure}[htbp]
    \centering
    \caption{(Second order) Plotting Result of \emph{plot\_D()}}
    \includegraphics[width=0.75\textwidth]{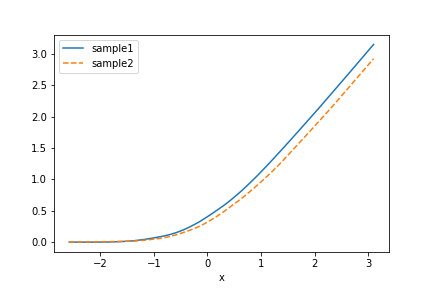}
    \label{fig:higher}
\end{figure}

Plotting a higher-order CDF can be conducted by
\begin{CodeChunk}
\begin{CodeInput}
>>> testing_normal_2 = pysdtest.test_sd(sample1, sample2, 
ngrid = 100, s = 2, resampling = 'bootstrap')
>>> testing_normal_2.plot_CDF(save=True, title="Normal distributions2.png") 
\end{CodeInput}
\end{CodeChunk}
Figure \ref{fig:higher} shows the corresponding result.

PySDTest provides three resampling procedures: bootstrap, paired bootstrap, and subsampling. The \textsf{subsampling()} function generates and returns subsamples $\mathcal{S}_{i,b}$. The argument b means the length of block, and nsub reflects the number of resampled test statistics. In testing codes, we set nsub as $N-b+1$.

\begin{CodeChunk}
\begin{CodeInput}
pytestsd.subsampling(sample, b, nsub)
\end{CodeInput}
\end{CodeChunk}

The \textsf{bootstrap()} function generates and returns bootstrap samples. The argument b means the number of observations for one boostrapping, which is used to calculate test statistics. In testing codes, we set b$=N$. The nboot means the number of bootstrapping. 

\begin{CodeChunk}
\begin{CodeInput}
pytestsd.bootstrap(sample, b, nboot)
\end{CodeInput}
\end{CodeChunk}

The \textsf{paired\_bootstrap()} function conducts bootstrapping in a pair $(X_{1,i},X_{2,i})$, and the description of arguments is the same as \textsf{bootstrap()}.

\begin{CodeChunk}
\begin{CodeInput}
pytestsd.paired_bootstrap(sample1, sample2, b, nboot)
\end{CodeInput}
\end{CodeChunk}

PySDTest provides two functions that help users calculate the test statistics by themselves. The \textsf{set\_grid()} function returns the `ngrid' number of grid points which are equally divided from the minimum to the maximum value of samples. For instance,
\begin{CodeChunk}
\begin{CodeInput}
>>> # Concatenate samples
>>> samples = [sample1, sample2]
>>> # Set grid
>>> pytestsd.set_grid(samples, ngrid)
\end{CodeInput}
\end{CodeChunk}

The \textsf{CDF()} calculates the (integrated) empirical CDF of the sample. For example,

\begin{CodeChunk}
\begin{CodeInput}
>>> pytestsd.CDF(sample1, grid, s = 1)
\end{CodeInput}
\end{CodeChunk}

\subsubsection[Advanced Hypothesis Testing]{Advanced Hypothesis Testing}

Through PySDTest, the user can perform advanced hypothesis regarding SD. For example, consider the case when there are $K$ distributions. Then, the hypothesis of our interest can be  $H_{0}^{(s)}:$ There exists at least one distribution that $s$-th order stochastically dominates some of the other distributions. $H_{1}^{(s)}$ is the negation of $H_{0}^{(s)}$. This can be expressed as
\begin{align}
    &H_{0}^{(s)}: \min_{k \neq l} \big\{ D_{k,l}^{(s)}(x)  \big\} \leq 0 \text{ for all } x \in \mathcal{X} \\
    &H_{1}^{(s)}: \min_{k \neq l} \big\{ D_{k,l}^{(s)}(x)  \big\} > 0 \text{ for some } x \in \mathcal{X}
\end{align}

We describe testing procedures step-by-step and conduct testing via PySDTest. Before testing, we generate random samples.

\begin{CodeChunk}
\begin{CodeInput}
>>> # Parameters
>>> mu1, mu2, mu3 = 0, 0.5, 1
>>> sigma1, sigma2, sigma3 = 1, 1.5, 2
>>> 
>>> # Sample size
>>> N = 1000
>>> 
>>> # set the seed
>>> np.random.seed(0)
>>>
>>> # Random numbers from a normal distribution
>>> sample1 = mu1 + sigma1 * np.random.randn(N)
>>> sample2 = mu2 + sigma2 * np.random.randn(N)
>>> sample3 = mu3 + sigma3 * np.random.randn(N)
\end{CodeInput}
\end{CodeChunk}

First, we need to set options for the testing; the user should specify the SD order, the number of grid points, and grid. In this example, we test first order SD and set the number of grid points as 100.
\begin{CodeChunk}
\begin{CodeInput}
>>> " Step 0: Setting Options "
>>> # SD order
>>> s     = 1
>>> # Setting grid
>>> samples = [sample1, sample2, sample3]
>>> ngrid = 100
>>> grid  = pysdtest.set_grid(samples, ngrid)
>>> # Setting the Number of Bootstrapping
>>> nsamp = 200
\end{CodeInput}
\end{CodeChunk}

Then, the user has to compute test statistics. This can be done by using the \textsf{CDF()} function in PySDTest. We consider the supremum-type test statistics, $T_{N}$, which may be written as
\begin{equation*}
    T_{N} = \sqrt{N}\min_{k \neq l} \big\{ \sup_{x \in \mathcal{X}} D_{k,l}^{(s)}(x)  \big\} \text{ for all } k,l \in \{1,2,...,K \}
\end{equation*}

For simplicity, we let $K=3$. Then, the sample analogue of the test statistics can be computed as:

\begin{CodeChunk}
\begin{CodeInput}
>>> " Step 1: Compute Test Statistics "
>>> # Function caculating CDF
>>> CDF = pysdtest.CDF
>>> 
>>> # Calculating D
>>> D_12 = CDF(sample1, grid, s) - CDF(sample2, grid, s)
>>> D_21 = CDF(sample2, grid, s) - CDF(sample1, grid, s)
>>> D_23 = CDF(sample2, grid, s) - CDF(sample3, grid, s)
>>> D_32 = CDF(sample3, grid, s) - CDF(sample2, grid, s)
>>> D_13 = CDF(sample1, grid, s) - CDF(sample3, grid, s)
>>> D_31 = CDF(sample3, grid, s) - CDF(sample1, grid, s)
>>> 
>>> # Calculating Test Statistics (In here, Supremum-type test statistics)
>>> D_collection = [D_12,D_21,D_23,D_32,D_13,D_31]
>>> test_stat = np.sqrt(N)*np.min(np.max(D_collection,1))
\end{CodeInput}
\end{CodeChunk}

Then we conduct resampling. In here, we generate bootstrap samples as

\begin{CodeChunk}
\begin{CodeInput}
>>> " Step 2: Generate Bootstrap Sample (or Subsample) "
>>> # Resampling (bootstrap)
>>> btspsample1 = pysdtest.bootstrap(sample1, b = N, nboot = nsamp)
>>> btspsample2 = pysdtest.bootstrap(sample2, b = N, nboot = nsamp)
>>> btspsample3 = pysdtest.bootstrap(sample3, b = N, nboot = nsamp)
\end{CodeInput}
\end{CodeChunk}

The next step is to calculate bootstrap test statistics and attain bootstrap distribution. We conduct recentered bootstrap as

\begin{CodeChunk}
\begin{CodeInput}
>>> " Step 3: Compute Test Statistics by Bootstrap Sample (or subsample) "
>>> # Calculating D by bootstrap samples
>>> D_b_12 = CDF(btspsample1, grid, s) - CDF(btspsample2, grid, s)
>>> D_b_21 = CDF(btspsample2, grid, s) - CDF(btspsample1, grid, s)
>>> D_b_23 = CDF(btspsample2, grid, s) - CDF(btspsample3, grid, s)
>>> D_b_32 = CDF(btspsample3, grid, s) - CDF(btspsample2, grid, s)
>>> D_b_13 = CDF(btspsample1, grid, s) - CDF(btspsample3, grid, s)
>>> D_b_31 = CDF(btspsample3, grid, s) - CDF(btspsample1, grid, s)
>>>             
>>> # Calculating Test Stat by bootstrap samples
>>> D_b_collection = [D_b_12,D_b_21, D_b_23, D_b_32, D_b_13, D_b_31]
>>> D_b_recentered = np.array(D_b_collection) - np.array(D_collection)
>>> resampled_stat = np.sqrt(N) * np.min(np.max(D_b_recentered,1),0)
\end{CodeInput}
\end{CodeChunk}

Then we compute the critical value under the significance level $\alpha$. Let $\alpha=0.05$, and

\begin{CodeChunk}
\begin{CodeInput}
>>> " Step 4: Compute the Critical Value of the Bootstrap Distribution "
>>> # Critical value and P-value (alpha = 0.05)
>>> alpha = 0.05
>>> critical_value = np.quantile(resampled_stat, 1 - alpha)
>>> pval = (resampled_stat >= test_stat).mean(0)
\end{CodeInput}
\end{CodeChunk}

Lastly, we decide whether to reject the null hypothesis or not. If the computed test statistics is larger than the critical value, reject $H_{0}^{(s)}$.

\begin{CodeChunk}
\begin{CodeInput}
>>> " Step 5: Reject H0 if Test stat > Critical Value "
>>> print("Test Statistic:   %6.3f" %  test_stat)
>>> print("Critical value:   %6.3f" % critical_value)
>>> print("P-value:          %6.3f" % pval)
\end{CodeInput}
\begin{CodeOutput}
Test Statistic:    0.791
Critical value:    0.506
P-value:           0.000
\end{CodeOutput}
\end{CodeChunk}

\subsection{Stata Command: \texttt{pysdtest}}

In this section, we introduce the Stata command \texttt{pysdtest}. The command is based on the python package PySDTest. Therefore, Stata with version $\geq$ 16.0, Python 3 software and installation of the Python package are necessary for using the command. Supported testing procedures include, but are not limited to, tests through the least favorable case approach of BD, subsampling approach of LMW, contact set approach of LSW and selective recentering method of DH. In addition, the NDM can be used to compute the critical value. \texttt{pysdtest} also supports the combination of various resampling methods, test statistics and procedures for approximating the limiting distribution.

Our hypothesis of interest is \eqref{eq:h0}; for given stochastic order $s$, the null hypothesis $H_0^s$ is that $X_1$ (Sample 1) $s$-th order stochastic dominates $X_2$ (Sample 2) and the alternative hypothesis $H_1^s$ is the negation of $H_0^s$. \texttt{pysdtest} allows arbitrary $s$-th SD order for testing. 

\subsubsection{Syntax and Options}

\begin{stsyntax}
pysdtest
\varlist(max = 2)\
\optif\
\optional{, options}
\end{stsyntax}

\texttt{varlist} specifies up to two variables for comparison. If only one variable is specified, a \texttt{by()} option must be used to divide the sample. The command supports several options to customize the testing procedure:
\begin{table}[htbp]
\centering
\resizebox{\textwidth}{!}{%
\begin{tabular}{lp{0.9\textwidth}}
\toprule
\textit{options} & \textbf{Description} \\ \midrule \\
\texttt{by(\varname)} & Specify a binary variable for dividing the sample. Without this, two variables are required for testing. \\ \\
\texttt{\underline{sw}itch} & Whether to switch the order of the sample when using \texttt{by()}. \\ \\
\texttt{resampling(\ststring)} & Resampling method (\textit{bootstrap}, \textit{subsampling}, or \textit{paired\_bootstrap}). Default is \textit{bootstrap}. For \textit{subsampling}, specify \texttt{b1()} and \texttt{b2()}. \\ \\
\texttt{approach(\ststring)} & Define a testing method. Default is the Kolmogorov-Smirnov type statistic as in BD and LMW. Alternatives include \textit{contact}, \textit{SR} (selective recentering), and \textit{NDM} (numerical delta method). \\ \\
\texttt{\underline{f}unctional(\ststring)} & For \textit{NDM}, specify a functional type (\textit{KS}, \textit{L1}, \textit{L2}); default is \textit{L1}. \\ \\
\texttt{ngrid(\num)} & Number of grid points for test statistics; default is 100. \\ \\
\texttt{s}(\textit{integer}) & Stochastic order of the null hypothesis; default is 1. \\ \\
\texttt{b1}(\textit{integer}) & Number of subsamples for the first variable. \\ \\
\texttt{b2}(\textit{integer}) & Number of subsamples for the second variable. \\ \\
\texttt{nboot}(\textit{integer}) & Number of bootstrapping iterations. \\ \\
\texttt{c(\num)} & Tuning parameter for the contact set approach; default is 0.75. \\ \\
\texttt{a(\num)} & Tuning parameter for the selective recentering approach; default is 0.1. \\ \\
\texttt{eta(\num)} & Tuning parameter for the selective recentering approach; default is $10^{-6}$. \\ \\
\texttt{epsilon(\num)} & Tuning parameter for the numerical delta method; default is 0.75. \\ \\
\texttt{alpha(\num)} & Significance level for the statistical test; default is 0.05. \\ \\ \bottomrule
\end{tabular}
}
\end{table}

\subsubsection{Stored Results}

The following scalars, macros and matrices are stored as a result of running \texttt{pysdtest}:

\begin{stresults}
\stresultsgroup{Scalars} \\
\stcmd{r(N1)} & \# of observations of the first variable
&
\stcmd{r(N2)} & \# of observations of the second variable
\\
\stcmd{r(b1)} & \# of subsamples of the first variable
&
\stcmd{r(b2)} & \# of subsamples of the second variable
\\
\stcmd{r(s)} & stochastic order
&
\stcmd{r(alpha)} & significance level
\\
\stcmd{r(ngrid)} & \# of grid points
&
\stcmd{r(test\_stat)} & value of the test statistic
\\
\stcmd{r(p\_val)} & p-value
&
\stcmd{r(critic\_val)} & critical value
\\ \\
\stresultsgroup{Macros} \\
\stcmd{r(resampling)} & resampling method
&
\stcmd{r(approach)} & testing approach
\\
\stcmd{r(form)} & type of functional for the NDM
\\ \\
\stresultsgroup{Matrices} \\
\stcmd{r(grid)} & matrix of grid values
&
\stcmd{r(limit\_dist)} & matrix of the limit distribution of test statistics approximated by resampling
\\
\end{stresults}

\subsubsection{Examples}

In this section, we describe example usages of \texttt{pysdtest} in Stata. We first generate the simulated data from the standard normal distribution.

\begin{stlog}
. 
. * Simulate the data
. * Clear the existing dataset
. clear
{\smallskip}
. 
. * Set the number of observations
. set obs 100
Number of observations ({\bftt{_N}}) was 0, now 100.
{\smallskip}
. 
. * Set the seed
. set seed 2024
{\smallskip}
. 
. * Generate a normally distributed variable
. gen X1 = rnormal(0, 1)
{\smallskip}
. gen X2 = rnormal(0, 1)
{\smallskip}
. 
 \nullskip
\end{stlog}

If there is no option given, \texttt{pysdtest} implements Kolmogorov-Smirnov type test statistics as in BD and LMW. The default resampling method is bootstrap and recentering as in BD.

\begin{stlog}
. 
. * Run pysdtest 
. pysdtest X1 X2
Running PySDTest
Sample1: X1
Sample2: X2
{\smallskip}
\#--- Testing for Stochastic Dominance  -----\#
{\smallskip}
* H0 : sample1 first order SD sample2
{\smallskip}
\#-------------------------------------------\#
{\smallskip}
*** Test Setting ***
* Resampling method      = bootstrap
* SD order               =      1
* \# of (sample1)         =    100 
* \# of (sample2)         =    100
* \# of bootstrapping     =    200
* \# of grid points       =    100
{\smallskip}
\#-------------------------------------------\#
{\smallskip}
*** Test Result ***
* Test statistic         = 0.2121
* Significance level     =  0.05
* Critical-value         = 1.0607
* P-value                = 0.8600
* Time elapsed           =  0.13 Sec
{\smallskip}
. 
\nullskip
\end{stlog}

The subsampling method as in LMW can be used by giving \textit{``subsampling"} as an argument to the option \texttt{resampling( )}. In this case, the user needs to set the number of subsamples by \texttt{b1( )} and \texttt{b2( )} options.

\begin{stlog}
. 
. * Run pysdtest with subsampling 
. pysdtest X1 X2, resampling("subsampling") b1(40) b2(40)
Running PySDTest
Sample1: X1
Sample2: X2
{\smallskip}
\#--- Testing for Stochastic Dominance  -----\#
{\smallskip}
* H0 : sample1 first order SD sample2
{\smallskip}
\#-------------------------------------------\#
{\smallskip}
*** Test Setting ***
* Resampling method      = subsampling
* SD order               =      1
* \# of (sample1)         =    100 
* \# of (sample2)         =    100
* \# of (subsample1)      =     40
* \# of (subsample2)      =     40
{\smallskip}
\#-------------------------------------------\#
{\smallskip}
*** Test Result ***
* Test statistic         = 0.2121
* Significance level     =  0.05
* Critical-value         = 1.0062
* P-value                = 0.9180
* Time elapsed           =  0.02 Sec
{\smallskip}
. 
\nullskip
\end{stlog}

In addition, users can employ the paired bootstrap method by giving \textit{``paired\_bootstrap"} as an argument to the option \texttt{resampling( )}. The number of observations for both sample1 and sample2 must be the same. 

Also, the contact set approach can be utilized by giving \textit{``contact"} to \texttt{approach( )} option. The additional tuning parameter can be chosen by \texttt{c( )} option. For the contact set approach, the integral type test statistic is employed as in \eqref{eq:lsw-test-stat}.

\begin{stlog}
. 
. * Run pysdtest with paired bootstrapping and contact set approach
. pysdtest X1 X2, resampling("paired_bootstrap") approach("contact") c(0.5)
Running PySDTest
Sample1: X1
Sample2: X2
{\smallskip}
\#--- Testing for Stochastic Dominance  -----\#
{\smallskip}
* H0 : sample1 first order SD sample2
* Contact Set Approach
{\smallskip}
\#-------------------------------------------\#
{\smallskip}
*** Test Setting ***
* Resampling method      = paired_bootstrap
* SD order               =      1
* \# of (sample1)         =    100 
* \# of (sample2)         =    100
* \# of bootstrapping     =    200
* \# of grid points       =    100
{\smallskip}
\# Tuning parameter -------
* c                      = 0.5000
{\smallskip}
\#-------------------------------------------\#
{\smallskip}
*** Test Result ***
* Test statistic         = 0.3800
* Significance level     =  0.05
* Critical-value         = 22.5270
* P-value                = 0.8400
* Time elapsed           =  0.13 Sec
{\smallskip}
. 
\nullskip
\end{stlog}

\texttt{pysdtest} provides \texttt{by( )} option to compare distributions divdided by a binary variable. As an illustration, we use automobile data stored in Stata and compare price distributions by the foreign dummy. 

\begin{stlog}
. 
. sysuse auto, clear
(1978 automobile data)
{\smallskip}
. 
. gen     foreign_str = "Domestic" if foreign == 0
(22 missing values generated)
{\smallskip}
. replace foreign_str = "Foreign"  if foreign == 1
(22 real changes made)
{\smallskip}
. 
. * Run pysdtest with by( ) option
. pysdtest price, by(foreign_str)
Running PySDTest
Groups:
`"Domestic"' `"Foreign"'
Sample1: price of Foreign
Sample2: price of Domestic
{\smallskip}
\#--- Testing for Stochastic Dominance  -----\#
{\smallskip}
* H0 : sample1 first order SD sample2
{\smallskip}
\#-------------------------------------------\#
{\smallskip}
*** Test Setting ***
* Resampling method      = bootstrap
* SD order               =      1
* \# of (sample1)         =     22 
* \# of (sample2)         =     52
* \# of bootstrapping     =    200
* \# of grid points       =    100
{\smallskip}
\#-------------------------------------------\#
{\smallskip}
*** Test Result ***
* Test statistic         = 0.3025
* Significance level     =  0.05
* Critical-value         = 1.1215
* P-value                = 0.7650
* Time elapsed           =  0.05 Sec
{\smallskip}
. 
\nullskip
\end{stlog}

If users want to switch the order of samples, they can use the option \texttt{switch} to swap the order of samples.

\begin{stlog}
. 
. * Run pysdtest with by( ) option and switch
. pysdtest price, by(foreign_str) switch
Running PySDTest
Groups:
`"Domestic"' `"Foreign"'
Sample1: price of Domestic
Sample2: price of Foreign
{\smallskip}
\#--- Testing for Stochastic Dominance  -----\#
{\smallskip}
* H0 : sample1 first order SD sample2
{\smallskip}
\#-------------------------------------------\#
{\smallskip}
*** Test Setting ***
* Resampling method      = bootstrap
* SD order               =      1
* \# of (sample1)         =     52 
* \# of (sample2)         =     22
* \# of bootstrapping     =    200
* \# of grid points       =    100
{\smallskip}
\#-------------------------------------------\#
{\smallskip}
*** Test Result ***
* Test statistic         = 0.8867
* Significance level     =  0.05
* Critical-value         = 1.0184
* P-value                = 0.1150
* Time elapsed           =  0.05 Sec
{\smallskip}
. 
\nullskip
\end{stlog}

NDM can be used by giving \textit{``NDM"} to \texttt{approach( )} option. We employ Kolmogorov-Smirnov type test statistic in this example. We can also set the number of grid points by \texttt{ngrid( )} option.

\begin{stlog}
. 
. * Run pysdtest with by( ) option, NDM
. pysdtest price, by(foreign_str) switch ///
> approach("NDM") functional("KS") ngrid(200)
Running PySDTest
Groups:
`"Domestic"' `"Foreign"'
Sample1: price of Domestic
Sample2: price of Foreign
{\smallskip}
\#--- Testing for Stochastic Dominance  -----\#
{\smallskip}
* H0 : sample1 first order SD sample2
* Numerical Delta Method
* KS Type Test Statistic 
{\smallskip}
\#-------------------------------------------\#
{\smallskip}
*** Test Setting ***
* Resampling method      = bootstrap
* SD order               =      1
* \# of (sample1)         =     52 
* \# of (sample2)         =     22
* \# of bootstrapping     =    200
* \# of grid points       =    200
{\smallskip}
\# Tuning paremeter -------------
* epsilon                = 0.9180
{\smallskip}
\#-------------------------------------------\#
{\smallskip}
*** Test Result ***
* Test statistic         = 0.9623
* Significance level     =  0.05
* Critical-value         = 1.0114
* P-value                = 0.0600
* Time elapsed           =  0.06 Sec
{\smallskip}
. 
\nullskip
\end{stlog}

\section{Empirical Illustration}

In this section, we compare historical daily returns of Bitcoin price and S$\&$P 500 index and investigate SD relations between them. This paper applies PySDTest to illustrate how portfolio comparison based on SD criteria can be done. SD concept is useful for portfolio comparison because SD provides a general rule for the portfolio selection under certain expected utility paradigm. We can interpret testing results in terms of preference to prospects such as monotonicity and risk aversion of an economic agent. The replication code is shown in Appendix.

In this paper, daily returns, $R_{t}^{k}$, are defined as:

\begin{equation}
    R_{t}^{k} = \log(\frac{P_{t}^{k}}{P_{t-1}^{k}}) \text{ for } t = 1,...,N \text{ and } k \in \{\text{Bitcoin}, \text{ S\&P 500} \}
\end{equation} where $P_{t}^{k}$ is price of Bitcoin or S$\&$P 500 index at day $t$.\footnote{We use closing price of Bitcoin and S$\&$P 500. Because the cryptocurrency market does not actually `close', we alternatively use price at 11:59 p.m. provided by the \href{https://coinmarketcap.com/}{CoinMarketCap}} Table \ref{table:desc-stat} summarizes the time length, number of observations and descriptive statistics of daily return data. Because the cryptocurrency market operates in weekends and holidays, the number of observations of Bitcoin price is more than S\&P 500 index. 

\begin{table}[htbp]
        \centering
        \caption{Descriptive Statistics of Bitcoin price and S$\&$P 500}
        \resizebox{\textwidth}{!}{
        \begin{tabular}{lcccccc}
        \toprule
        Name & Time & Observations & Mean & Std & Min & Max \\
        \midrule
        Bitcoin & Apr.29.2013 - Feb.27.2021 & 2861 & 0.0020 & 0.0426 & -0.4647 & 0.3575 \\ 
        S$\&$P 500 & Apr.29.2013 - Feb.27.2021 & 1995 & 0.0004 & 0.0109 & -0.1277 & 0.0897 \\
\bottomrule
\end{tabular}}
\label{table:desc-stat}
\end{table}

\begin{figure}[htbp]
    \centering
    \caption{CDF of Bitcoin and S\&P 500 Index}
    \includegraphics[width = 0.9\textwidth]{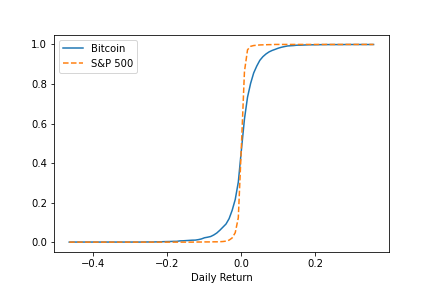}
    \caption*{\footnotesize{\emph{Notes} This figure shows cumulative distribution functions of daily returns of Bitcoin price and S\&P 500 index. The red dash line indicates the distribution function of daily return of Bitcoin. The blue solid line indicates that of S\&P 500 index.}}
    \label{fig:crypto-cdf}
\end{figure}

Standard mean-variance analysis cannot determine superiority of daily returns between Bitcoin and S\&P 500 index. Although the mean return of Bitcoin price is larger than that of S$\&$P 500 index, standard deviation of Bitcoin is also larger. In addition, the distribution of Bitcoin price seems to be more dispersed than that of S$\&$P 500 index. Figure \ref{fig:crypto-cdf} shows cumulative distribution functions of daily return of Bitcoin and S$\&$P 500 index.\footnote{This is plotted by using the function \emph{plot\_CDF()} in PySDTest}

We conduct testing for the following null hypotheses to investigate SD relations:
\begin{align}
    H_{0,1}^{s}&: R^{\text{Bitcoin}} \succeq_{s} R^{\text{S\&P 500}} \\
    H_{0,2}^{s}&: R^{\text{S\&P 500}} \succeq_{s} R^{\text{Bitcoin}}
\end{align} for $s=1,2$ The alternative hypotheses $H_{1,i}^{s}$ are the negation of $H_{0,i}^{s}$ for $i,s=1,2$, respectively. We use the subsampling method by LMW in order to account for time-series dependence. We set 100 grid points. 1,000 and 900 are set as a size of subsamples for daily returns of Bitcoin and S\&P 500 index, respectively. Table \ref{table:crpto-test-result}
 summarizes the testing results.

\begin{table}[htbp]
        \centering
        \caption{Testing Results}
        \resizebox{0.75\textwidth}{!}{
        \begin{tabular}{lcccccc}
        \toprule
        $H_{0}$ &  &\multicolumn{2}{c}{$R^{\text{Bitcoin}} \succeq_{s} R^{\text{S\&P 500}}$} & & \multicolumn{2}{c}{$R^{\text{S\&P 500}} \succeq_{s} R^{\text{Bitcoin}}$} \\
        \cmidrule{3-4} \cmidrule{6-7}  
        $s$    &   & 1 & 2 & & 1 & 2  \\
        \midrule  
        p-value &  & 0.0000 & 0.0000 &  & 0.0000 & 0.3349 \\
\bottomrule
\end{tabular}}
\caption*{\footnotesize{ \emph{Notes.} We use the subsampling method and set 100 grid points. 1,000 and 900  subsamples are set for daily return of Bitcoin and S\&P 500 index, respectively.}}
\label{table:crpto-test-result}
\end{table}

In Table \ref{table:crpto-test-result}, all hypotheses of the first SD order ($s=1$) are rejected. This means that an economic agent with monotone increasing utility function strictly prefer neither Bitcoin nor S\&P 500 to the other. In contrast, it is shown that daily returns of S\&P 500 second order ($s=2$) stochastically dominate those of Bitcoin. This implies that an agent with risk-averse utility function prefers investing in S\&P 500 index to Bitcoin. 

\section{Conclusion}

In this paper, we have introduced a Python package, PySDTest, and a Stata command, \texttt{pysdtest}. The package implements various Stochastic Dominance (SD) tests, including those proposed by BD, LMW, LSW, and DH. PySDTest includes testing using the NDM, which applies modern developments in statistical inference methods to directionally differentiable functions. We provide guidance for using PySDTest and \texttt{pysdtest}, including testing for extended notions of the SD hypothesis. We applied PySDTest to compare the daily return distributions between Bitcoin prices and the S\&P 500 index. The results show that the S\&P 500 second-order stochastically dominates Bitcoin returns.

Empirical implementation of the SD concept has not been as extensive as it could be despite the fundamental role of the SD concept in diverse research areas and the recent advances in inference procedures. This may be partly due to the lack of publicly available software. In this paper, we provide a comprehensive Python package and Stata command for SD tests to assist empirical practitioners and academics. We look forward to seeing many applications of our software that help practitioners as intended.

\newpage

\bibliographystyle{aer}
\bibliography{ref.bib}

\newpage

\section*{A. Appendix}

\subsection{Replication Code for Section 4.}
\subsubsection{Replication by Python codes}
\begin{lstlisting}[language = Python]
""" import modules """

import pandas as pd
import numpy as np
import pysdtest
import matplotlib.pyplot as plt

""" Descriptive Analysis """

BTC_daily_rr  = pd.read_excel("BTC_daily_rr.xlsx")
sp_daily_rr   = pd.read_excel("sp_daily_rr.xlsx")

BTC_daily_rr.describe()
sp_daily_rr.describe()

# Datatype for tests should 1-dim numpy array
BTC_daily_rr = np.array(BTC_daily_rr['d_ln_Close'])
sp_daily_rr = np.array(sp_daily_rr['d_ln_Close'])

""" Plotting """

# Plotting D
crypto_testing_ct_1 = pysdtest.test_sd(BTC_daily_rr, sp_daily_rr, ngrid = 100, s=  1, b1 = 1000, b2 = 900, resampling =  'subsampling')
crypto_testing_ct_1.plot_CDF(save=True, title="CDFs of Bitcoin and S&P Index.png",label1="Bitcoin",label2="S&P 500", xlabel = "Daily Return")

""" Testing H0: BTC >= S&P500 """

# First order

crypto_testing_1 = pysdtest.test_sd(BTC_daily_rr, sp_daily_rr, ngrid = 100, s=  1, b1 = 1000, b2 = 900, resampling =  'subsampling')
crypto_testing_1.testing()

# Second order

crypto_testing_2 = pysdtest.test_sd(BTC_daily_rr, sp_daily_rr, ngrid = 100, s= 2, b1 = 1000, b2 = 900, resampling = 'subsampling')
crypto_testing_2.testing()

""" Switching samples 

Testing H0 : S&P500  >= BTC

"""
# First order

crypto_testing_sw_1 = pysdtest.test_sd(sp_daily_rr, BTC_daily_rr, ngrid = 100, s = 1, b1 = 900, b2 = 1000, resampling = 'subsampling')
crypto_testing_sw_1.testing()

# Second order

crypto_testing_sw_2 = pysdtest.test_sd(sp_daily_rr, BTC_daily_rr, ngrid = 100, s= 2, b1 = 900, b2 = 1000, resampling = 'subsampling')
crypto_testing_sw_2.testing()
\end{lstlisting}

\subsubsection{Replication by the Stata Command}

\begin{stlog}
. clear
{\smallskip}
. use bitcoin_sp500_daily_rr, replace
{\smallskip}
. 
. * Specify the python environment
. set python_exec /opt/anaconda3/bin/python
{\smallskip}
. python query // Check the python version
\HLI{168}
    Python Settings
      set python_exec      /opt/anaconda3/bin/python
      set python_userpath  
{\smallskip}
    Python system information
      initialized          no
      version              3.11.8
      architecture         64-bit
      library path         /opt/anaconda3/lib/libpython3.11.dylib
{\smallskip}
. 
. * Run
. pysdtest BTC_daily_rr SP500_daily_rr, ///
>  resampling("subsampling") s(1) b1(1000) b2(900)
Running PySDTest
Sample1: BTC_daily_rr
Sample2: SP500_daily_rr
{\smallskip}
\#--- Testing for Stochastic Dominance  -----\#
{\smallskip}
* H0 : sample1 first order SD sample2
{\smallskip}
\#-------------------------------------------\#
{\smallskip}
*** Test Setting ***
* Resampling method      = subsampling
* SD order               =      1
* \# of (sample1)         =   2861 
* \# of (sample2)         =   1995
* \# of (subsample1)      =   1000
* \# of (subsample2)      =    900
{\smallskip}
\#-------------------------------------------\#
{\smallskip}
*** Test Result ***
* Test statistic         = 6.1859
* Significance level     =  0.05
* Critical-value         = 4.0054
* P-value                = 0.0000
* Time elapsed           =  6.83 Sec
{\smallskip}
. pysdtest BTC_daily_rr SP500_daily_rr, ///
>  resampling("subsampling") s(2) b1(1000) b2(900)
Running PySDTest
Sample1: BTC_daily_rr
Sample2: SP500_daily_rr
{\smallskip}
\#--- Testing for Stochastic Dominance  -----\#
{\smallskip}
* H0 : sample1 second order SD sample2
{\smallskip}
\#-------------------------------------------\#
{\smallskip}
*** Test Setting ***
* Resampling method      = subsampling
* SD order               =      2
* \# of (sample1)         =   2861 
* \# of (sample2)         =   1995
* \# of (subsample1)      =   1000
* \# of (subsample2)      =    900
{\smallskip}
\#-------------------------------------------\#
{\smallskip}
*** Test Result ***
* Test statistic         = 0.3177
* Significance level     =  0.05
* Critical-value         = 0.2147
* P-value                = 0.0000
* Time elapsed           =  6.88 Sec
{\smallskip}
. 
. pysdtest SP500_daily_rr BTC_daily_rr, ///
>  resampling("subsampling") s(1) b1(900) b2(1000)
Running PySDTest
Sample1: SP500_daily_rr
Sample2: BTC_daily_rr
{\smallskip}
\#--- Testing for Stochastic Dominance  -----\#
{\smallskip}
* H0 : sample1 first order SD sample2
{\smallskip}
\#-------------------------------------------\#
{\smallskip}
*** Test Setting ***
* Resampling method      = subsampling
* SD order               =      1
* \# of (sample1)         =   1995 
* \# of (sample2)         =   2861
* \# of (subsample1)      =    900
* \# of (subsample2)      =   1000
{\smallskip}
\#-------------------------------------------\#
{\smallskip}
*** Test Result ***
* Test statistic         = 8.2384
* Significance level     =  0.05
* Critical-value         = 6.7111
* P-value                = 0.0000
* Time elapsed           =  6.80 Sec
{\smallskip}
. pysdtest SP500_daily_rr BTC_daily_rr, ///
>  resampling("subsampling") s(2) b1(900) b2(1000)
Running PySDTest
Sample1: SP500_daily_rr
Sample2: BTC_daily_rr
{\smallskip}
\#--- Testing for Stochastic Dominance  -----\#
{\smallskip}
* H0 : sample1 second order SD sample2
{\smallskip}
\#-------------------------------------------\#
{\smallskip}
*** Test Setting ***
* Resampling method      = subsampling
* SD order               =      2
* \# of (sample1)         =   1995 
* \# of (sample2)         =   2861
* \# of (subsample1)      =    900
* \# of (subsample2)      =   1000
{\smallskip}
\#-------------------------------------------\#
{\smallskip}
*** Test Result ***
* Test statistic         = 0.0541
* Significance level     =  0.05
* Critical-value         = 0.0783
* P-value                = 0.3349
* Time elapsed           =  6.88 Sec
{\smallskip}
. 
\nullskip
\end{stlog}

\end{document}